\documentclass[twocolumn]{aastex7}
\usepackage{draftwatermark}
\SetWatermarkScale{3.5}
\SetWatermarkColor[gray]{0.88}
\usepackage{xcolor}
\usepackage{gensymb}
\usepackage{subfigure}
\usepackage{tabularx}
\usepackage{tablefootnote}
\usepackage{amsmath} 
\usepackage{hyperref} 
\shorttitle{Extra-Tidal Stars: NGC 6569}
\shortauthors{Hughes et al.}

\newcommand\bes{Besan\c con}
\newcommand\spc{SP\_Ace}
\newcommand\afe{\ensuremath{[\alpha/\mathrm{Fe}]_{\rm SP}}}
\newcommand\fesp{\ensuremath{\mathrm{[Fe/H]}_{\rm SP}}}
\newcommand\fedp{\ensuremath{\mathrm{[Fe/H]}_{\rm DP}}}
\newcommand\kms{km~s$^{-1}$}
\usepackage{xcolor} 

\newcommand{\rev}[1]{\textcolor{black}{#1}}

\begin{document}

\title{The Milky Way Bulge Extra-Tidal Star Survey: \\ NGC 6569}

 \author{Joanne Hughes}
\email{jhughes@seattleu.edu}
 \affiliation{Physics Department, Seattle University, 901 12th Ave., Seattle, WA 98122, USA}
 \author{Andrea Kunder}
\email{AKunder@stmartin.edu}
 \affiliation{Saint Martin's University, 5000 Abbey Way SE, Lacey, WA 98503, USA}
 \author{Kevin Covey}
\email{coveyk@wwu.edu}
 \affiliation{Department of Physics \& Astronomy, Western Washington University, MS-9164, 516 High St., Bellingham, WA 98225, USA}
 \author{Kathryn Devine}
\email{KDevine@collegeofidaho.edu}
 \affiliation{The College of Idaho, 2112 Cleveland Blvd Caldwell, ID 83605, USA}
 \author{Kristen A. Larson}
\email{krislars@gmail.com}
 \affiliation{National Solar Observatory, 3665 Discovery Dr, Boulder, CO 80303, USA}
 \author{Carlos Campos}
\email{c.a.campos9177@gmail.com}
 \affiliation{Saint Martin's University, 5000 Abbey Way SE, Lacey, WA 98503, USA}
 \author{Adrian M. Price-Whelan}
\email{adrianmpw@gmail.com}
 \affiliation{Center for Computational Astrophysics, Flatiron Institute, 162 Fifth Ave., New York,  NY 10010, USA}
 \author{Joseph E. McEwen}
\email{jmcewen@seattleu.edu}
 \affiliation{Physics Department, Seattle University, 901 12th Ave., Seattle, WA 98122, USA}
 \author{Gabriel I. Perren}
\email{gabrielperren@gmail.com}
 \affiliation{Instituto de F\'isica de Rosario, IFIR (CONICET-UNR), 2000 Rosario, Argentina}
 \author{Christian I. Johnson}
\email{chjohnson1@stsci.edu}
 \affiliation{Space Telescope Science Institute, 3700 San Martin Drive, Baltimore, MD 21218, USA}
 \author{Craig Horton}
\email{chorton@seattleu.edu}
 \affiliation{Physics Department, Seattle University, 901 12th Ave., Seattle, WA 98122, USA}
 \author{Luke Smith}
\email{lsmith4@seattleu.edu}
 \affiliation{Physics Department, Seattle University, 901 12th Ave., Seattle, WA 98122, USA}
 \author{Sarah Torset}
\email{torsetsarah@seattleu.edu}
 \affiliation{Physics Department, Seattle University, 901 12th Ave., Seattle, WA 98122, USA}
 \author{Cynthia Luna}
\email{cyn.luna02@gmail.com}
 \affiliation{Physics Department, Seattle University, 901 12th Ave., Seattle, WA 98122, USA}
 \author{Matthew Kolmanovsky}
\email{kolmanovskym@seattleu.edu}
 \affiliation{Physics Department, Seattle University, 901 12th Ave., Seattle, WA 98122, USA}
 \author{Fiona Kovisto}
\email{fkovisto@seattleu.edu}
 \affiliation{Physics Department, Seattle University, 901 12th Ave., Seattle, WA 98122, USA}
 \author{Leander Villarta}
\email{lvillarta@seattleu.edu}
 \affiliation{Physics Department, Seattle University, 901 12th Ave., Seattle, WA 98122, USA}
 \author{Vy Vuong}
\email{vvuong@seattleu.edu}
 \affiliation{Computer Science Department, Seattle University, 901 12th Ave., Seattle, WA 98122, USA}
 \author{Iulia T. Simion}
\email{iuliateodorasim@yahoo.com}
 \affiliation{Shanghai Key Lab for Astrophysics, Shanghai Normal University, 100 Guilin Road, Shanghai, 200234, People’s Republic of China}
 \author{Kyle Webster}
\email{webstek6@wwu.edu}
 \affiliation{Department of Physics \& Astronomy, Western Washington University, MS-9164, 516 High St., Bellingham, WA 98225, USA}
 \author{Erika Silva}
\email{silvae3@wwu.edu}
 \affiliation{Fernbank Science Center, 156 Heaton Park Dr, Atlanta, GA 30307, USA}
 \author{Catherine A. Pilachowski}
\email{cpilacho@iu.edu}
 \affiliation{Indiana University, Department of Astronomy, SW319, 727 E 3rd Street, Bloomington, IN 47405, USA}
 \author{R. Michael Rich}
\email{rmrastro@gmail.com}
 \affiliation{Department of Physics and Astronomy, UCLA, 430 Portola Plaza, Box 951547, Los Angeles, CA 90095-1547, USA}
 \author{Justin A. Kader}
\email{jukader@iu.edu}
 \affiliation{Department of Physics and Astronomy, 4129 Frederick Reines Hall, University of California, Irvine, CA 92697, USA}
 \author{Andreas J. Koch-Hansen}
\email{andreas.koch@uni-heidelberg.de}
 \affiliation{Zentrum f\"ur Astronomie der Universit\"at Heidelberg, Astronomisches Rechen-Institut, M\"onchhofstr. 12, 69120 Heidelberg, Germany}
 \author{Meridith Joyce}
\email{meridith.joyce@gmail.com}
 \affiliation{Department of Physics and Astronomy, University of Wyoming, 1000 E University Ave, Laramie, WY 82071, USA}
 \author{Sean McAdam}
\email{mcadams@wwu.edu}
 \affiliation{Department of Physics \& Astronomy, Western Washington University, MS-9164, 516 High St., Bellingham, WA 98225, USA}
 \author{Faith Benda}
\email{bendaf@wwu.edu}
 \affiliation{Department of Physics \& Astronomy, Western Washington University, MS-9164, 516 High St., Bellingham, WA 98225, USA}




\begin{abstract}
We present spectroscopic evidence for tidal debris associated with the bulge
globular cluster NGC~6569, based on medium-resolution ($R\sim 11{,}000$)
spectra of 303 stars obtained with the Anglo-Australian Telescope. Targets were
selected using Blanco DECam Bulge Survey (BDBS) photometry with \textit{Gaia}
DR3 astrometry, and span $\sim 7\arcmin$-$30\arcmin$ (i.e., $1$-$5\,r_t$, where
$r_t$ is the King-model tidal radius) from the cluster center. Theoretical
modeling shows that the Jacobi radii can vary between $8\arcmin$-$11\arcmin$
and $18\arcmin$-$22\arcmin$ over the orbit, likely leaving stars around the
cluster that are transitioning into the predicted leading and lagging tidal
tails. We identify 40 stars in this sample that exhibit chemical and kinematic
properties consistent with previous, or borderline, cluster membership. The seven
best candidates have ${\rm S/N}>30$, with
${\rm [Fe/H]} = -0.83 \pm 0.14$~dex and [$\alpha$/Fe] $= +0.38 \pm 0.06$~dex,
consistent with NGC~6569's bound population. Our findings provide evidence that
NGC~6569 is actively losing stars through tidal stripping, contributing to the
bulge field population at a present rate of
$1.0$-$1.6\,M_\odot~{\rm Myr^{-1}}$, which corresponds to
$\approx 5.6 \pm 1.3\%$ of its present-day mass per Gyr. This work is part of
the Milky Way Bulge Extra-Tidal Star Survey (MWBest) and represents our first
detailed study of a massive bulge globular cluster in this context.
 \end{abstract}

 

\section{Introduction}

Globular clusters (GCs) are key probes of the Milky Way’s formation.
Their chemical compositions and orbital dynamics offer critical insights into
early galaxy assembly, the Galactic potential, accretion history, and bulge
enrichment \citep[e.g.,][]{hozumi15, deboer19, piatti20b}.

NGC~6569 is a relatively massive
\citep[$M \approx 2.3\times 10^5 M_\odot$;][]{baumgardt21}, moderately
metal-rich bulge cluster with
$-0.9 \lesssim [\mathrm{Fe/H}] \lesssim -0.7$~dex
\citep[e.g.,][]{valenti11, johnson18, crociati23, barrera25}.
It has been extensively studied via high-resolution spectroscopy,
\textit{Gaia} DR3 astrometry, and deep photometry
\citep[e.g.,][]{hazen85, ortolani01, valenti05, valenti11, kunder15,
johnson18, saracino19, pallanca23}.  NGC~6569 exhibits two red clumps
along its horizontal branch (HB), separated by $\sim 0.1$~mag in the $K_S$
band, as revealed by VVV photometry \citep{minniti10, mauro12}.
\citet[][hereafter J18]{johnson18} used Magellan-M2FS and VLT-FLAMES
spectra ($R \sim 27{,}000$) to analyze RGB and HB stars, finding that both
HB groups share the cluster’s motion (albeit consistent with the bulge field
as well) and differ in metallicity by only $0.13$~dex, a statistically
insignificant offset given the small sample.
J18 reported a mean radial velocity of $-48.8 \pm 5.3$~\kms\ and an average
metallicity of ${\rm [Fe/H]} = -0.87 \pm 0.05$~dex from 19 M2FS spectra,
along with 100 CaT metallicities from FLAMES spectra,
${\rm [Fe/H]}_{\rm CaT} = -0.84 \pm 0.17$~dex (used here as
${\rm [Fe/H]}_{\rm J18}$).  \citet[][hereafter P23]{pallanca23} and \citet{crociati23} combined
multi-instrument ESO/VLT spectroscopy, \textit{Gaia} DR3 proper motions,
and star counts to fit King models and trace member stars out to
$r \sim 770\arcsec$. P23 conducted a detailed analysis of the internal
kinematics and structure of NGC~6569, showing that its velocity-dispersion
and density profiles are well reproduced by a single-mass King model and
finding only a weak hint of ordered rotation in an intermediate radial
range. However, both J18 and P23 targeted RGB stars at radii extending
beyond the commonly used $r_t \approx 6\farcm9$ value \citep{valenti11},
which helped motivate our spectroscopic search for extra-tidal debris around NGC~6569 in 2022, where the results were first presented in a \citet[AAS poster]{hughes23}, ahead of the discussion in \citet{barrera25}. This latest study analyzed APOGEE-2 $H$-band spectra of 11 RGB members with BACCHUS, deriving
${\rm [Fe/H]} = -0.91 \pm 0.06$~dex,
$[\alpha/{\rm Fe}] = +0.36 \pm 0.06$~dex, and
$RV = -49.8 \pm 3.7$~{\kms}, and detecting multiple populations via a large
N-spread and a clear C-N anticorrelation \citep{barrera25}.

Theoretical models suggest that bulge GCs within $\sim 2$~kpc of the Galactic
center may lose up to $80\%$ of their mass to the field \citep{baumgardt18},
potentially contributing to the bulge’s double red clump feature
\citep{lim21}.  In the halo, \citet{piatti20b} found that $\sim 26\%$ of GCs
exhibit tidal tails, $42\%$ show distinct extra-tidal features, and $32\%$
show no detectable structure.  In contrast to halo clusters like Pal~5
\citep{odenkirchen01, odenkirchen03, grillmair06, erkal17}, tidal features
are rarely detected around bulge GCs, with only a few examples such as
NGC~6355 and NGC~6362 \citep{zhang22, piatti24a, piatti24b}.

This study is part of the Milky Way Bulge Extra-Tidal Star Survey (MWBest),
an effort to understand globular cluster dissolution in the inner Galaxy.
MWBest targets are selected using photometric data from the Blanco DECam
Bulge Survey (BDBS) \citep{rich20, johnson20} and \textit{Gaia} DR3 astrometry
\citep{gaia22, marton23}.  BDBS provides deep $ugrizY$ photometry over
$200$~deg$^2$ of the southern bulge using DECam; these passbands are very
similar to those of the Vera Rubin Observatory, whose bulge observations
will be incorporated into future MWBest work.  Spectroscopic observations
use the AAOmega spectrograph on the 3.9~m Anglo-Australian Telescope, with
the red 1700D grating centered at 8600\AA\ ($R \sim 11{,}000$; range:
8350-8800\AA).

In this paper we (1) present spectroscopic confirmation of extra-tidal
stars around NGC~6569 using an 8-dimensional chemo-dynamical selection
(PM, RV, [Fe/H], [$\alpha$/Fe]) with an explicit Monte Carlo bulge-contamination
test; (2) measure the present-day mass-loss rate using both a central
“snapshot” estimator and a tail-based, time-averaged estimator; and
(3) propagate the dominant systematic uncertainties (orbit sampling in the
rotating bar, spectroscopic selection completeness $f_{\rm sel}$, and
tracer-to-mass conversion $\langle m\rangle$) into $\dot M / M_{\rm cl}$.
The outline of the paper is as follows: \S2 summarizes prior measurements
(Table~1) and our photometric and spectroscopic data (Tables~2-4),
\S3 compares the observations with theoretical models, \S4 discusses the
results, \S5 summarizes our conclusions, and the Appendix examines the
extra-tidal candidates from an energy and escape-velocity standpoint and
quantifies the variations in the Jacobi radius along their bulge-confined
orbits.

\section{Observations and Data Reduction} 

\subsection{Physical Properties of NGC~6569}
NGC 6569 is located at $RA = 18^h 13^m 38.3^s$ and $Dec = -31^\circ 49^\prime 35^{\prime\prime}$. Table~1 provides a summary of the relevant previous work, and Figure~1 shows a pictorial description of the cluster environment and local \textit{Gaia} DR3 PMs. The cluster is estimated to be 13.0 Gyr old with $[\mathrm{Fe/H}] = -0.75$ dex, fit with the MIST isochrones \citep[and this paper]{choi16, saracino19}. We generated custom MIST isochrones that take $\alpha$-element enhancement into account in the prescription of the metallicity. We also used the Automated Stellar Cluster Analysis code  \citep [ASteCA]{perren15} to fit cluster properties using Bayesian statistics. Distance estimates range from 12.4~kpc \citep{barrera25}, through 10.9 kpc \citep{harris96,harris10,valenti11} to 10.1~kpc (P23). The \citet{baumgardt21} catalog (2021; Version~4 as of 2023) places the cluster at $10.53 \pm 0.26$~kpc using \textit{Gaia} DR3 data and estimates its mass as $2.3 \times 10^5 M_\odot$ based on N-body simulations; we use these catalog values for the rest of the analysis in this paper.

\begin{figure*}
\centering
\includegraphics[height=4.0in, width=1.0\textwidth]{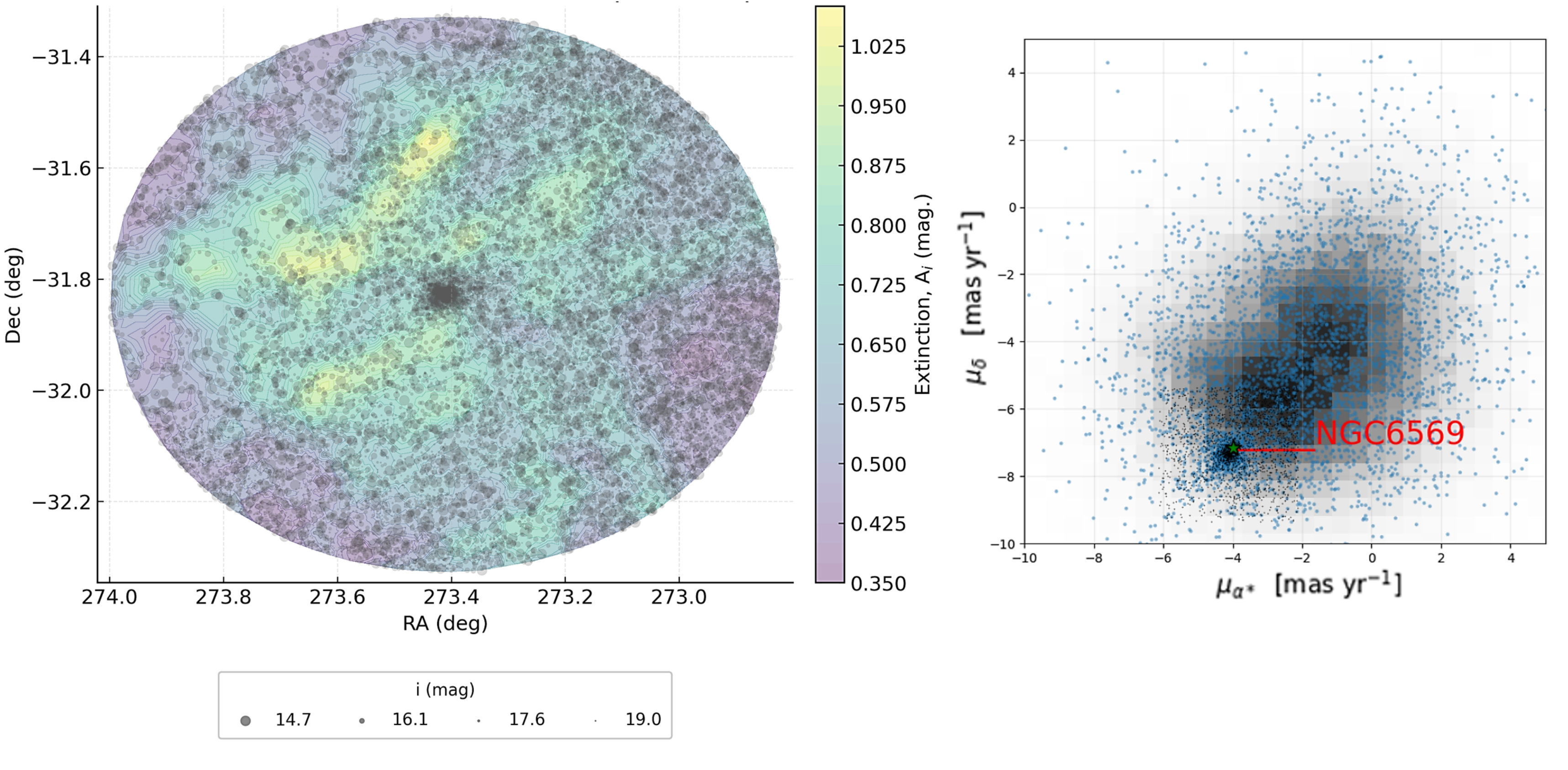}
\caption{\textbf{(a)} The 18,180 objects within a radius of 30\arcmin\ of the cluster center, at $RA=273.\degree412000$, $Dec=-31.\degree826444$  with BDBS {\it i}-band measurements between 10.4 and 19.8 mag., overlaid on a variable extinction map \citep{simion17,kader23} in the BDBS {\it i}-band. \textbf{(b)}  Gaia~DR3 sources were queried from \texttt{gaiadr3.gaia\_source} around the SIMBAD center of NGC\,6569 and filtered to have measured proper motions, $G<19$, and $\mathrm{RUWE}<1.4$. Three subsets are shown: (i) an inner $5\arcmin$ circle around the cluster center (``cluster''; blue points), (ii) a $15\arcmin$-$60\arcmin$ annulus (``bulge/field''; gray density histogram, randomly sub-sampled to $1.2\times10^{5}$ rows via \texttt{random\_index} for speed), and (iii) a $2\arcmin$ core used only to estimate the cluster mean motion. The green star marks the median cluster proper motion from the $2\arcmin$ core. Tiny black dots show proper motions from our BDBS catalog overplotted on the same axes. Axes are $\mu_{\alpha*}\equiv \mathrm{d}\alpha/\mathrm{d}t \cos\delta$ versus $\mu_{\delta}$, in $\mathrm{mas\,yr^{-1}}$. No parallax or CMD cleaning is applied here, so the $5\arcmin$ (blue) selection includes some field contaminants.}
\end{figure*}

\subsection{Photometric Analysis}
The ASteCA code \citep{perren15} was applied to BDBS photometry, combined with \textit{Gaia} astrometry, to estimate the overall physical properties of NGC~6569 and assess membership probabilities for stars within the cluster radius ($r_{cl}$). This radius is not the tidal radius, but rather the point at which the stellar density surrounding the cluster fades into the background field. ASteCA employs a decontamination algorithm that identifies and rejects stars from surrounding fields, selecting likely members based on their location within the cluster radius.

The code computes key structural properties of the cluster, including its center, stellar density profile, luminosity function, completeness limits, and color-magnitude diagram (CMD). Cluster parameters were derived using three-parameter King models \citep{king62, king66}, with the resulting values listed in Table~1. Photometry was dereddened using extinction maps from \citet{simion17}. CMD fitting was performed using solar-scaled PARSEC isochrones \citep{bressan12}, which yield total metallicities [M/H] that are typically higher than the corresponding [Fe/H].

It is important to note that King models do not account for mass segregation or the dynamical evolution of globular clusters. As clusters evolve, they lose mass via stellar evolution and dynamical interactions, which can reshape their density profiles. Such mass loss often results in an extended outer profile that, when fit with a King model, can overestimate the cluster’s true size and mass \citep{baumgardt17}. To address these limitations, ASteCA has expanded its capabilities to include alternative profiles, such as Plummer models, as demonstrated in recent applications \citep{rain24}. These enhanced modeling tools will be employed in our future analysis of this system. 

\subsubsection{Contamination and Membership Estimation}

To evaluate the degree of field star contamination in the cluster region, we employed the ASteCA contamination index (CI), defined as:
\begin{equation}
    \mathrm{CI} = \frac{d_{\mathrm{field}}}{(n_{\mathrm{cl}} + n_{\mathrm{fl}})/A_{\mathrm{cl}}} = \frac{n_{\mathrm{fl}}}{n_{\mathrm{cl}} + n_{\mathrm{fl}}},
\end{equation}
where $d_{\mathrm{field}}$ is the field star density, $A_{\mathrm{cl}}$ is the area of the best-fit King model, and $n_{\mathrm{cl}}$ and $n_{\mathrm{fl}}$ are the numbers of stars attributed to the cluster and field, respectively.

ASteCA returned a CI of 0.35 for stars within the adopted PM range, indicating that approximately 35\% of the sample within the accepted GC boundaries consists of field interlopers. 
For this GC, RVs and chemical abundances are necessary to find extra-tidal stars, as \textit{Gaia} PMs alone are not sufficient.

Above $G = 19$~mag, ASteCA identified $529 \pm 5$ members based on the King profile, and $538 \pm 5$ when assuming a uniform field background. Its Bayesian decontamination algorithm uses photometric properties to assign membership probabilities and provide statistical uncertainties: capabilities that traditional isochrone fitting lacks. However, if tidal debris lies beyond the modeled cluster radius, these stars may be erroneously excluded.

\subsubsection{Comparison with Prior Studies and Systematic Effects}

Table~1 (Notes 4 to 9) summarizes the impact of varying search radius, extinction corrections, and PM filters on derived structural parameters. In the crowded and dusty bulge environment, differential extinction and high stellar background complicate surface brightness fitting, potentially introducing systematic errors \citep{piatti20b}.

ASteCA’s distance estimate agrees with the N-body results from the Galactic Globular Cluster Database (Version 4; \citealt{baumgardt21}), though the catalog's mass estimate may be more robust due to their simulation-based approach. As shown in Table~1, King model tidal radius estimates range from $r_t \sim$5\farcm5 (ASteCA) to 9\farcm8 (P23), depending on contamination handling: over-subtraction of field stars yields artificially small $r_t$, while under-subtraction inflates it. Although we filtered stars outside the cluster’s PM envelope before modeling, some contamination persisted, as reflected in the CI value.

Due to distance and patchy extinction, our photometry does not reach the main-sequence turnoff. If NGC~6569 is mass-segregated, low-mass stars would preferentially reside in the cluster outskirts-regions that may be undersampled. We used the King tidal radius of 6\farcm9 \citep{ortolani01,valenti11} to define the cluster-field boundary when observing runs were planned in 2022 (P23 was published after the spectra were obtained).

\begin{deluxetable*}{ccccccccl}
\setlength{\tabcolsep}{4pt}   
\tablewidth{0pt}              
\tablecaption{NGC 6569: Physical Parameters \& Model Fits \label{tab:deluxetable}}
\tablehead{\\
 \colhead{Distance} & 
 \colhead{Mass} & 
 \colhead{$\log_{10}$(Age)} & 
 \colhead{Z or [Fe/H]} & 
 \colhead{E(B-V)} & 
 \colhead{$r_{\mathrm{cl}}$} & 
 \colhead{$r_t$} & 
 \colhead{RV} & 
 \colhead{Note} \\
 \colhead{(kpc)} & 
 \colhead{($10^5 M_\odot$)} & 
 \colhead{(dex)} & 
 \colhead{(dex)} & 
 \colhead{(mag)} & 
 \colhead{(\arcmin)} & 
 \colhead{(\arcmin)} & 
 \colhead{(\kms)} &
}
\startdata 
 $10.53\pm 0.26$ & $2.3$ & 10.08 & Z=0.008 & 0.49 & 2.7 & 6.9 & $-49.82\pm 0.50$ & [1] \\
 $9.8 - 12.05$ & $"$ & $-$ & $[\mathrm{Fe/H}]=-0.79\pm 0.02,\;[\alpha/\mathrm{Fe}]=+0.4\pm 0.02$ & $-$ & $-$ & $6.\arcmin9 -7.\arcmin15$ & $-47\pm 4$ & [2] \\
 $10.1\pm 0.2$ & $1.72^{+0.20}_{-0.18}$ & $-$ & $[\mathrm{Fe/H}]=-0.90\pm 0.24$ & $-$ & $-$ & $9.83^{+2.8}_{-1.8}$ & $-50.5\pm 0.3$ & [3] \\
 $10.50\pm 0.27$ & $-$ & $-$ & $-$ & $-$ & $3.28^{+0.17}_{-0.20}$ & $4.56^{+1.23}_{-0.53}$ & $-$ & [4] \\
 $10.50\pm 0.27$ & $-$ & $-$ & $-$ & $-$ & $3.95^{+0.62}_{-0.68}$ & $5.45^{+0.63}_{-0.50}$ & $-$ & [5] \\
 $10.8\pm 0.3$ & $4.9\pm 0.4^*$ & $10.02\pm 0.02$ & $Z=0.0081\pm 0.0005$ & $0.49\pm 0.03$ & $5.2^{+0.6}_{-0.5}$ & $15.1^{+0.4}_{-0.3}$ & $-$ & [6] \\
 $10.2\pm 0.3$ & $2.4\pm 0.1^*$ & $10.00\pm 0.02$ & $Z=0.0081\pm 0.0003$ & $0.027\pm 0.003^{**}$ & $3.92\pm 0.01$ & $5.50\pm 0.04$ & $-$ & [7] \\
 $10.5\pm 0.3$ & $-$ & 10.08 & $[\mathrm{Fe/H}]=-0.75,\;[\alpha/\mathrm{Fe}]=+0.3$ & \textit{Variable} & $-$ & $-$ & $-48.8\pm 5.3$ & [8] \\
 $10.5\pm 0.3$ & $1.8\pm 1.3^*$ & $10.00\pm 0.04$ & $Z=0.0078\pm 0.0020$ & $0.47\pm 0.10$ & $10.7^{+1.4}_{-3.6}$ & $15.1^{+10.9}_{-8.2}$ & $-44.1\pm 11.3$ & [9] \\
 $12.4\pm 1.45$& $-$ &$-$ & $[\mathrm{Fe/H}]=-0.91\pm 0.06,\;[\alpha/\mathrm{Fe}]=+0.36\pm 0.06$ & 0.68& $-$ & $-$ & $-49.75 \pm 3.68$ & [10] \\
\enddata
\tablecaption{Cluster Notes \label{tab:notes}}
\tablecomments{\\
\textsuperscript{[1]} Catalog compilation: \citet{vasiliev21, baumgardt23, baumgardt21, baumgardt18, valenti11}-Galactic GC Database V4 (2023). \\
\textsuperscript{[2]} \citet{ortolani01, valenti05, harris10, valenti11}. \\
\textsuperscript{[3]} \citet{pallanca23, crociati23}; CaT-based metallicities. \\
\textsuperscript{[4]} ASteCA King model fit: \textit{Gaia} Plx, PM, and photometry, $r=0.5^\circ$, PM-restricted, no reddening correction \citep{perren15, king62, king66}. \\
\textsuperscript{[5]} ASteCA King fit: BDBS+Gaia-matched photometry, $r=0.5^\circ$, PM-restricted, no reddening correction. \\
\textsuperscript{[6]} ASteCA fit: \textit{Gaia} photometry, $r=1^\circ$, PM-restricted, no reddening correction; PARSEC isochrones \citep{bressan12}. \\
\textsuperscript{[7]} ASteCA fit: BDBS+Gaia photometry, $r=0.5^\circ$, $Plx < 0.4$ mas, PM-restricted, reddening corrected. \\
\textsuperscript{[8]} Empirical CMD fit with extinction map correction and MIST isochrones \citep{choi16, simion17, kader23}. \\
\textsuperscript{[9]} ASteCA fit: \textit{Gaia} photometry, $r=1^\circ$, \textit{Gaia} RV and PM restricted, no reddening correction. \\
\textsuperscript{[10]} \citet{barrera25}.\\
\textsuperscript{*} Mass estimates using the \citet{kroupa02} IMF; lower limits due to the omission of dynamical mass loss \citep{pera22}.\\
\textsuperscript{**} Residual extinction after correction using the \citet{simion17} parametric extinction map.
}
\end{deluxetable*}

\subsection{Spectroscopic Target Selection}

To account for differential reddening across the bulge field (Figure~1a), we applied extinction corrections using spatially resolved maps from \citet{simion17}, based on the $R_V = 3.1$ and the \citet{fitzpatrick99} extinction law. The mean reddening was $E(B-V) = 0.37 \pm 0.07$~mag., with values ranging from 0.19 to 0.55~mag. (listed in Table~2). 
The commonly used extinction coefficients \citep{simion17, kader23} were:
\begin{align*}
   A_u &= 4.239\,E(B-V), & \; A_g &= 3.384\,E(B-V),\\ 
   \; A_r &= 2.483\,E(B-V), & \; A_i &= 1.838\,E(B-V). \end{align*}

Our spectroscopic target selection prioritized BDBS sources with reliable $u$- and $i$-band photometry. The inclusion of the {\it u}-band in BDBS photometry significantly enhances the identification of multiple stellar populations (MPs), as it probes near-UV features sensitive to CNO abundances. \citet{kader22} demonstrated that $u$-band photometry is especially effective for distinguishing chemically distinct populations in GCs. \textit{Gaia} photometry, in contrast, lacks coverage in these wavelength regions. 

To isolate extra-tidal candidates beyond our assumed cluster radius (King model $r_t \sim 6\farcm9$; \citealt{ortolani01, harris10, valenti11}), we began with proper motion cuts based on the cluster’s \textit{Gaia} DR3 kinematics: $-5.12 < \mu_\alpha \cos(\delta) < -3.13$~mas~yr$^{-1}$ and $-8.85 < \mu_\delta < -5.85$~mas~yr$^{-1}$ (a rectangular cut), and incorporating a parallax cut of $Plx \leq 0.4$~mas to minimize foreground contamination. We summarize the process in Figure~2, with the red open squares indicating objects within the PM range. Further constraints were imposed: objects with BDBS {\it u}-magnitudes with uncertainties $<0.04\; mag.$  were selected if they had detections in the BDBS {\it g}- and {\it i}-bands. 

\begin{figure*}
\includegraphics[width=0.90\textwidth]{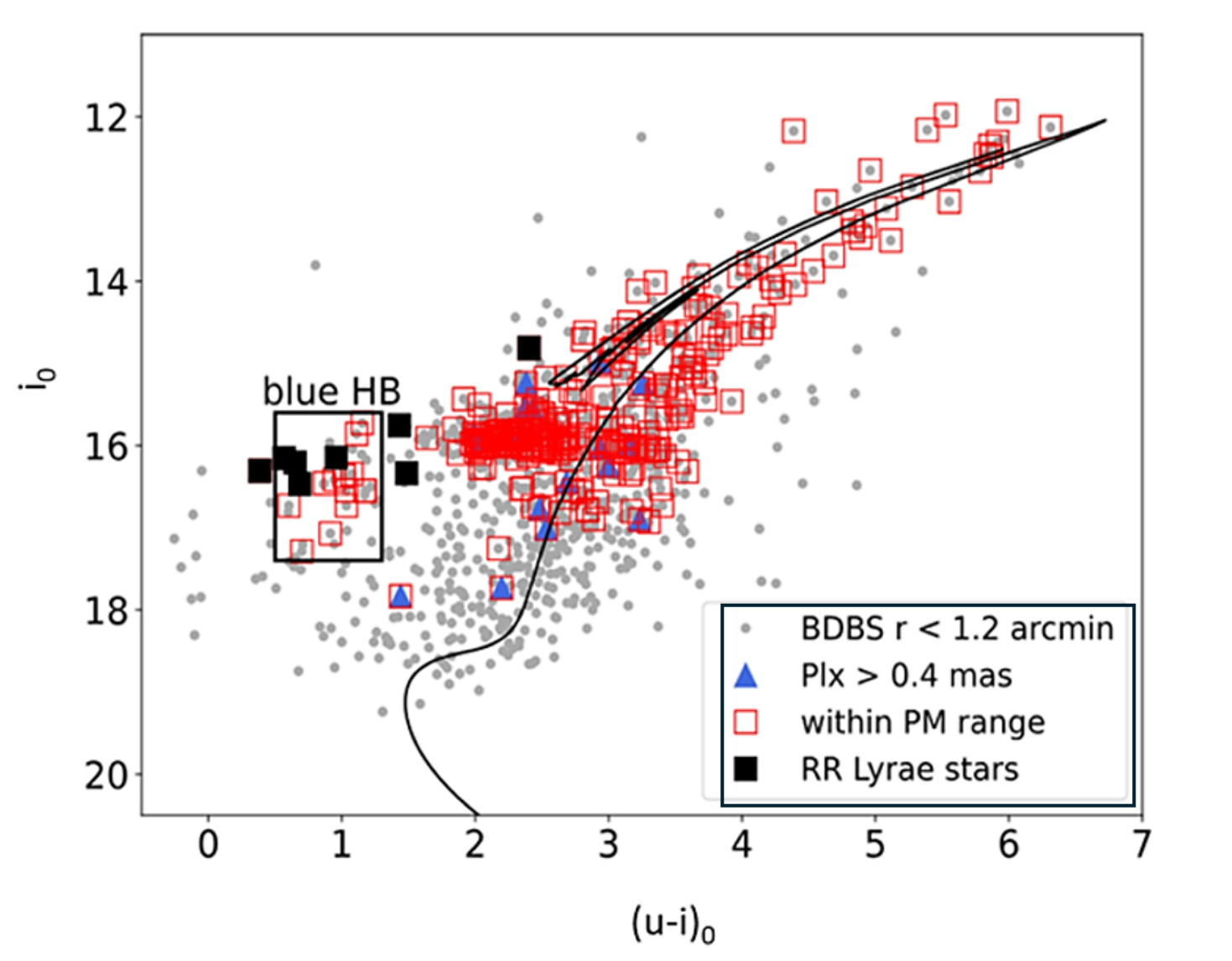}
\caption{The BDBS differentially-dereddened CMD for $i_0\; vs.\; (u-i)_0$ is shown. The small, gray-filled circles are the stars within 1\farcm2 of the center of NGC~6569. The following objects are identified within 5 tidal radii of the center of the cluster: stars within 0.95$^\circ$ of the cluster center; stars must have a PM within 1 mas$\,yr^{-1}$ in RA and within 1.5 mas$\,yr^{-1}$ in Dec of the main cluster PM: a box drawn between ($\sim -4.125 \pm 1.0$ mas$\,yr^{-1}$, $\sim -7.354 \pm 1.5$ mas$\,yr^{-1}$).\rm \ Additional constraints were applied: objects with BDBS u-magnitudes with uncertainties $<0.04\; mag.$  were selected if they had detections in the BDBS g- and {\it i}-bands also. The red open squares were identified as targets for AAT/AAOmega spectroscopy. Blue triangles are flagged as likely foreground objects, and black-filled squares are identified as RRLs. The isochrone shown is from the MIST models \citep{choi16}, $\textrm{[Fe/H]} = -0.75$~dex, scaled for $[ \alpha /Fe] = +0.3$~dex external to MIST, with an age of 13~Gyr. }
\end{figure*}

Although \textit{Gaia} DR3 provides extensive coverage, its spectroscopic completeness is limited in crowded bulge regions \citep[P23]{saracino19,brown21}. Of the 61,411 \textit{Gaia} sources within 1\degree\ of NGC~6569, only 3,997 have radial velocities, and just 401 lie beyond the Galactic Center. To distinguish potential cluster escapees from field stars, we adopted an expected radial velocity window of $-65 < \mathrm{RV} < -25$~km~s$^{-1}$. This range captures the observed dispersion of the cluster’s systemic velocity ($\mathrm{RV} \approx -49.9$~km~s$^{-1}$) and includes stars consistent with leading and trailing tidal debris (see \S3). It also avoids the peak of bulge contaminants, which dominate outside this range. This selection is consistent with previous MWBest studies \citep[e.g.,][]{kunder24,butler24}.

\subsection{Spectroscopic Observations}

 \citet{kunder24} details our usage of the AAOmega multifibre spectrograph at the 3.9-m Anglo-Australian Telescope (AAT, Siding Spring Observatory, Coonabarabran, NSW, Australia). The spectra taken around NGC~6569 were part of a five-night observing run between July 20$^{th}$ and  July 24$^{th}$, 2022 (PROP-ID: O/2022A/3002). The plate configurations for the Two Degree Field (2dF) fiber positioner contained a combination of RR Lyrae (RRL) stars (including four in the center of the cluster), RC stars, HB stars, and giants centered on the cluster, half-filling the 2-degree field of view. 
 \rev{The RRL targets in our study were selected from the fourth data release of the Optical Gravitational Lensing Experiment's (OGLE-IV) bulge RR~Lyrae catalogs \citep{udalski15,soszynski14,soszynski19}, which provide a large, well-characterized sample with multi-epoch photometry in the $V$ and $I$ passbands. Optical photometry for our full target set (and the CMD context for the RRL candidates) is taken from the BDBS catalogs \citep{rich20,johnson20}. This combination of OGLE-based RRL identification and BDBS photometric characterization follows the same strategy adopted in recent MWBest/BDBS cluster studies \citep[e.g.,][]{kunder24,butler24}.}

 The AAO 2dfdr pipeline \citep{aao15} was used to reduce the spectra: bias subtraction, cosmic ray cleaning, quartz flat-fielding, wavelength calibration via arc-lamp exposures, and sky subtraction using dedicated sky fibers. The spectra analyzed in this paper have a useful wavelength range of 8420-8800\AA , with slight variations depending on the exact position of the spectra on the CCD.

We employed the red 1700D grating (centered at 8600\AA ) to detect the CaT lines for all objects and the blue 2500V grating (centered at 5000\AA ) to detect the Mg line at 5180\AA\ for the brightest stars. However, this paper only reports the results derived from the red spectra, as the light cirrus and mediocre seeing during these observations prevented the signal-to-noise (S/N) ratio from being sufficient to utilize the blue spectra. Depending on weather and other factors, the exposure times ranged from $4\times 30$ minutes to $2\times 30$ minutes. It is possible to robot-position 392 fibers anywhere within the $2^\circ$ prime-focus field; each fiber core subtends about 2\arcsec\ on the sky. We positioned 305 fibers within 1-5 tidal radii of the center of NGC~6569, which yielded 303 readable spectra. We show an example spectrum, used for RV-measurement from the CaT-lines in Figure~3, and list the results in Table~3. The map of the fiber positions with the RV results is shown in Figure~4a.

\subsubsection{Radial Velocities} 

We elected to use the mean wavelength shift of the CaT lines to calculate each target stars' RV. The CaT lines are among the strongest features in the near-infrared spectra of late-type stars, making them relatively easy to detect even in faint objects \citep{grocholski06, carrera07}. This strength is advantageous when observing distant or low-luminosity stars, as it allows for accurate measurements of  RV and equivalent widths, which are crucial for finding stray stars with RVs consistent with the cluster and deriving metallicities. The prominence of these lines enables us to obtain reliable data from stars that might otherwise be too faint for detailed analysis using optical lines, which are often weaker and more susceptible to blending with other spectral features \citep{cole04}.

We compared the results of measuring the RVs with {\tt IRAF} and {\tt astropy}, and found them consistent, but the velocity uncertainties were smaller using the {\tt astropy} method, averaging the central wavelengths of the three strongest lines, the CaT. The barycentric correction was made using the {\tt PyAstronomy} routine to match the standard star results to the catalog standard frame of reference. We used {\tt astropy} routines to continuum-fit, normalize, find the CaT line centers, and measure their equivalent widths. For all readable spectra, the median and mean velocity errors were $\sim$2.4~\kms\ and 3.0~\kms . The hotter, fainter HB stars had a median and mean velocity error of $\sim$2.7 and 3.6~\kms . We report these data in Table~3. As with the BH~261 spectra \citep{kunder24}, RVs were also measured with IRAF's {\tt xcsao} routine \citep{kurtz92}. This process cross-correlates against spectra from three stars observed during the same run and used as templates, chosen from the Apache Point Observatory Galaxy Evolution Experiment \citep[APOGEE,][]{eisenstein11} database. \citet{kunder24} used three APOGEE stars as RV templates:
APOGEE 2M18134674\allowbreak-2926056 (RV=$27.88\pm0.03$~\kms),
APOGEE 2M17514997\allowbreak-2906055 (RV=$-187.33\pm0.02$~\kms), and
APOGEE 2M17521244\allowbreak-2919510 (RV=$65.13\pm0.05$~\kms),
yielding a median velocity error of $\sim3$~\kms\ for the upper RGB and
$9$~\kms\ for the fainter and hotter HB stars.

The positional map of the RV results is shown in Figure~4a\rm , with the RV color-coded circles showing all the 303 AAT AAOmega spectra, the triangles mark the local RRLs, and the green crosses are the J18 sample. Each successive gray circle denotes 1-5  tidal radii  \citep[ 6\farcm9]{ortolani01,valenti11}. Figure~4b \rm adds the J18 (bright green) sample to our AAT data (alone, the black lines with no shading), in relative frequencies. Comparisons are made between the source density histograms for NGC~6569's FOV (green) and the local \bes\ \citep{czekaj14} model population (mauve),  using the PM limits. The \bes\ models of the MW bulge are sophisticated representations designed to simulate the structure and stellar populations of the galaxy. The \bes\ models are consistent with the bulge field stars contaminating the cluster by about 35\%, as indicated by Equation (1). We describe our use of the \bes\ models in more detail in \S2.5, while assessing the statistical significance of extra-tidal star candidates compared to the characteristics of the local bulge population.

J18 reports that NGC~6569's bound population is confined to the range: $-30 > RV > -63$~kms$^{-1}$. Our velocity resolution is generally less than $\pm 3$~\kms , justifying the extension of the search range to $-25 > RV > -65$~kms$^{-1}$ within the AAT dataset. 
We center the RV-window on the systemic velocity from J18 and widen it modestly relative to J18 to reflect our lower spectral resolution; this captures possible tail members without excluding bona fide cluster stars. Because the cut is applied symmetrically about NGC~6569's $v_{\rm sys}$ and membership is evaluated in the full chemo-dynamical space, the choice does not bias us towards selecting potential escapees from the leading versus lagging tails.
We verified that the adopted window remains appropriate over our projected radii, given the declining dispersion profile, by checking that accepted candidates follow the modeled tail trend ${\rm RV}(s)$ on each side of the cluster.

\subsubsection{CaT Calibration}

Figure~3 \rm illustrates an example spectrum of one red giant branch (RGB) object (\#215 in Table~2), showing the CaT method of determining the RV from an average of the shifts of the 3 CaT lines, and also measuring their equivalent widths, after normalizing the spectrum. The RVs were shifted to the rest frame and the barycentric correction was applied from the telescope value to the velocity reported in Table~3, where we also report the measured CaT EWs for all 303 AAT spectra.

 The process of calibrating the equivalent widths (EWs) of the Calcium Triplet (CaT) lines to derive stellar metallicities, specifically the iron abundance [Fe/H], follows a multi-step procedure in this study. The CaT lines at 8498\AA , 8542\AA , and 8662\AA\ are measured, and their sum is defined as: \begin{equation} \Sigma EW = EW_{8498} + EW_{8542} + EW_{8662}. \end{equation}

To correct for luminosity effects, we applied the \citet{dias20} calibration using the DECam $i$-band magnitude, following the relation: \begin{equation} \Sigma EW = W^\prime_i - \beta_i (i - i_{\mathrm{HB}}), \end{equation} where $W^\prime_i$ is the reduced equivalent width, $\beta_i = 0.62 \pm 0.06 \pm 0.17$ is the slope of the luminosity correction (with the first uncertainty being formal and the second the standard deviation), and $i_{\mathrm{HB}} = 16.59$~mag. is the apparent {\it i}-band magnitude of the horizontal branch (HB), not corrected for reddening. The inclusion of reddening corrections altered the final [Fe/H] values by less than $\pm0.05$ dex, confirming its minor impact in this regime. The \citet{dias20,parisi22} calibrations for the CaT-method are matched to the methods established by \citet{cole04}, and based on the \citet{carretta97} scale. 

For comparison, we evaluated the calibration used by J18, which relied on the redder two CaT lines (8542\AA\  and 8662\AA ,) and the $K$-band magnitude. To estimate $\Sigma EW$ from two lines ($EW_{2L}$), we applied the empirical conversion: \begin{equation} \Sigma EW_{3L} = 1.26(\pm 0.13) + 1.00(\pm 0.03)\cdot EW_{2L}. \end{equation}

The final iron abundance was derived using the empirical linear calibration: \begin{equation} [\mathrm{Fe/H}]_{\mathrm{DP}} = 2.966(\pm 0.032) + 0.362(\pm 0.014) \cdot W^\prime, \end{equation} where \fedp\ denotes the metallicity derived via the \citet{dias20} method, anchored in CaT measurements. The results are reported in Table~4. Full CaT-to-[Fe/H] calibrations are only trustworthy for stars at NGC~6569's distance.

\begin{figure*}
\centering
\makebox[\textwidth][c]{\includegraphics[width=0.95\textwidth]{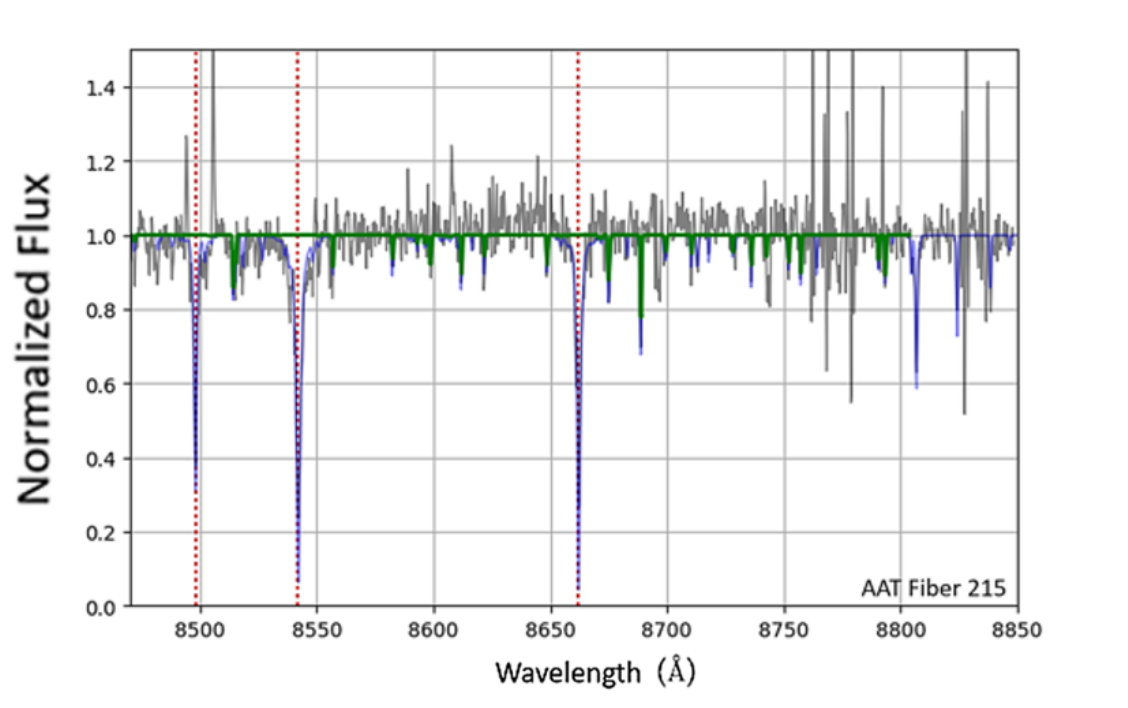}}
%
\caption{An example spectrum is shown with the regions used for the spectral analysis process identified. This is an extracted, processed, and normalized spectrum for AAT/AAOmega fiber \#215. The normalized spectrum is shown in gray. The centroid of each CaT feature is shown as a red dotted line, which has been shifted back to the rest frame. The {\tt specutils} routine was used to find lines by the derivative method, and the results were averaged for the three CaT lines. The blue line is the standard RGB spectrum which best fits the target star using the \spc\ code \citep{boeche16, boeche21}, and the green line is the fitted region, not using the calcium triplet, and avoiding the sky-lines at longer wavelengths than 8800\AA. The results for this star are: \fedp $=-0.92\pm 0.09$ dex, \fesp $=-0.83\pm 0.08$ dex, and  $[ \alpha/Fe] =+0.27\pm 0.27$~dex.}
\end{figure*}

\begin{figure*}[ht!]
\centering
\includegraphics[width=1.05\textwidth]{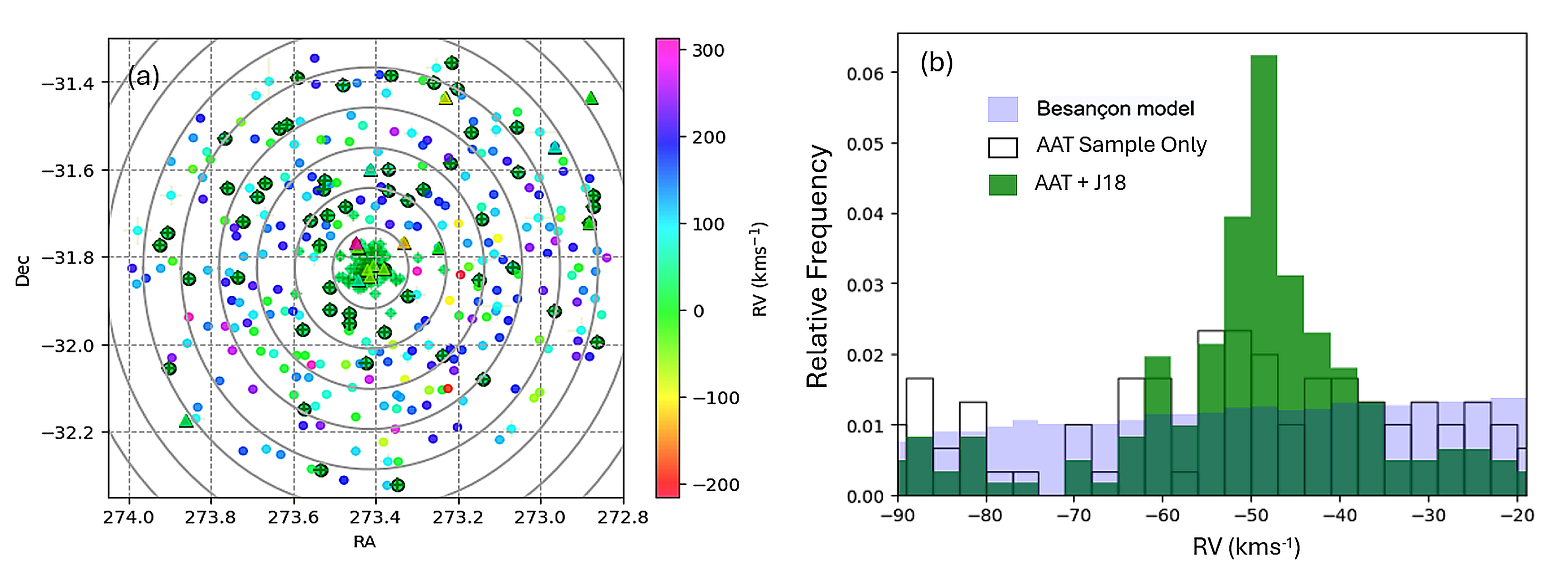}
\caption{\textbf{(a)} A map of the RA vs. Dec positions of the 303 spectroscopic targets (2 fibers did not yield measurable spectra) in and around NGC~6569 (shaded circles), along with some known RR Lyrae stars (OGLE) with cataloged RVs (filled triangles). The circles are color-coded by RV (see color bar on the right). The 100 objects from J18 are shown as green crosses in the cluster center. The gray circles are successive tidal radii \citep[6\farcm9 ]{ortolani01,valenti11} from the center of NGC~6569. We choose our search limit as $-25>RV>-65$~\kms , yielding 58 objects (black-edged green circles with black pluses). \textbf{(b)} The relative number of stars per bin is shown in this histogram of the cluster RV-range yielded by J18, from $-30>RV>-63$~\kms . The 100 stars from J18 were added to our 303 targets to find the cluster (green bars). The AAT sample only is shown as the black line (no shading). The \bes\ Galactic population models for our FOV and PM limits are seen in mauve. }
\end{figure*}

\begin{deluxetable*}{llllllllllllllll}
\tabletypesize{\tiny}
\tablewidth{0pt} 
\tablecaption{BDBS Data \& Extinction Corrections}
\tablehead{\\
 \colhead{No.} &  \colhead{RA(\degree )} & \colhead{Dec(\degree )} & 
 \colhead{u} & \colhead{$\sigma_u$}& \colhead{g} & \colhead{$\sigma_g$}& \colhead{r} &\colhead{$\sigma_r$}& \colhead{i} & \colhead{$\sigma_i$}&\colhead{z} & \colhead{$\sigma_z$}& \colhead{y} & \colhead{$\sigma_y$}&
 \colhead{E(B-V)} \\ 
 \colhead{(1)} & \colhead{(2)} & \colhead{(3)} & \colhead{(4)} & 
 \colhead{(5)} & \colhead{(6)}& \colhead{(7)} & \colhead{(8)}& \colhead{(9)} &\colhead{(10)}& \colhead{(11)} & \colhead{(12)}&\colhead{(13)} & \colhead{(14)}& \colhead{(15)} & \colhead{(16)}
\\}
\startdata 
1&  273.95715& -31.84990& 20.046& 0.014& 17.557& 0.004& 16.508& 0.009& 15.937& 0.004& 15.616& 0.003& 15.413& 0.007&  0.368 \\
2&  273.80745& -31.83143& 20.000& 0.039& 17.048& 0.001& 15.879& 0.008& 15.269& 0.002& 14.917& 0.011& 14.675& 0.011&  0.403\\
3&  273.99068& -31.82702& 19.640& 0.021& 16.908& 0.004& 15.841& 0.008& 15.290& 0.001& 14.962& 0.007& 14.757& 0.011& 0.318\\
 \multicolumn{16}{l}{\it Complete table available online.}
\enddata
\tablecomments{\tiny This table is available in its entirety in machine-readable form.\\
BDBS DECam filters and extinction corrections, as in \citet{simion17,kader23}.\\
(1) AAT AAOmega fiber no.\\
(2) \textit{Gaia} DR3 RA\\
(3) \textit{Gaia} DR3 Dec\\
(4) BDBS {\it u}-band mag.\\
(5) Uncertainty in BDBS {\it u}-band\\
(6) BDBS {\it g}-band mag.\\
(7) Uncertainty in BDBS {\it g}-band\\
(8) BDBS {\it r}-band mag.\\
(9) Uncertainty in BDBS {\it r}-band\\
(10) BDBS {\it i}-band mag.\\
(11) Uncertainty in BDBS {\it i}-band\\
(12) BDBS {\it z}-band mag.\\
(13) Uncertainty in BDBS {\it z}-band\\
(14) BDBS {\it y}-band mag.\\
(15) Uncertainty in BDBS {\it y}-band\\
(16) E(B-V) calculated according to \citet{simion17}, with the extinction law as discussed in \citet{kader23}, in mag.
}
\end{deluxetable*}

\begin{deluxetable*}{lllllllllllllll}
\tablewidth{0pt} 
\tablecaption{\textsc{Gaia} Data \& CaT Measurements}
\tablehead{ \\
\colhead{No.} & \colhead{RA} & \colhead{Dec} & \colhead{Plx} & \colhead{$\sigma_{Plx}$}& \colhead{$\mu_ \alpha cos(\delta)$} & \colhead{$\sigma_{\mu_ \alpha}$}& \colhead{$\mu_\delta$} & \colhead{$\sigma_{\mu_\delta}$}& \colhead{RV} & \colhead{$\sigma_{RV}$} &\colhead{S/N} &\colhead{$EW_{8498}$} & \colhead{$EW_{8542}$} & \colhead{$EW_{8662}$} \\
 \colhead{(1)} & \colhead{(2)} & \colhead{(3)} & \colhead{(4)} & 
 \colhead{(5)} & \colhead{(6)}& \colhead{(7)} & \colhead{(8)}& \colhead{(9)} &\colhead{(10)}& \colhead{(11)} & \colhead{(12)}&\colhead{(13)} & \colhead{(14)}& \colhead{(15)}   \\}
\startdata
1& 273.957165& -31.84991691& 0.0276& 0.0740& -3.2532& 0.0714& -5.9616& 0.0564& 91.8& 1.3& 16& 1.287& 3.39& 1.835\\
2& 273.8074592& -31.83144141& 0.1228& 0.0457& -3.2069& 0.0491& -8.3173& 0.0369& 35.5& 2.2& 16& 1.586& 4.174& 2.497\\
3& 273.9906913& -31.82703443& -0.0024& 0.0617& -4.3467& 0.0529& -8.2425& 0.0403& 137.4& 3.0& 31& 1.094& 3.225& 2.361\\
\multicolumn{15}{l}{\it Complete table available online.}
\enddata
\tablecomments{This table is available in its entirety in machine-readable form.\\
(1) AAT AAOmega fiber no.\\
(2) \textit{Gaia} DR3 RA\\
(3) \textit{Gaia} DR3 Dec\\
(4) \textit{Gaia} DR3 Parallax in {\it mas}\\
(5) Uncertainty in \textit{Gaia} DR3 Parallax in {\it mas}\\
(6) \textit{Gaia} DR3 PM in RA in {\it mas$\,yr^{-1}$}\\
(7) Uncertainty in \textit{Gaia} DR3 PM in RA in {\it mas$\,yr^{-1}$}\\
(8) \textit{Gaia} DR3 PM in Dec in {\it mas$\,yr^{-1}$}\\
(9) Uncertainty in \textit{Gaia} DR3 PM in Dec in {\it mas$\,yr^{-1}$}\\
(10) RV in \kms \\
(11) Uncertainty in RV in \kms \\
(12) SNR of spectrum \\
(13) EW of CaT1 in  \AA \\
(14) EW of CaT2 in  \AA \\
(15) EW of CaT3 in  \AA \\
}
\end{deluxetable*}

\subsubsection{Abundance Analysis with \spc}

In addition to using CaT lines, we employed \spc \ (Stellar Parameters and Chemical Abundances Estimator) to calculate both [M/H] and [Fe/H]. \spc\ \citep{boeche16, boeche21} is an open-source spectral synthesis tool designed to derive stellar parameters and chemical abundances. It supports resolving powers from $R = 2{,}000$ to $40{,}000$, which encompasses our AAT spectra at $R \sim 11{,}000$. Originally developed for the RAVE survey \citep{kunder17, steinmetz20}, \spc\ is capable of measuring individual abundances of Mg, Al, Si, Ca, Ti, Fe, and Ni for our AAT spectra with $S/N > 35$, provided both red and blue spectral coverage is available. Tests found for our red AAT spectra, only [Fe/H] is reliably determined across the full dataset. We designate these metallicity measurements as \fesp\ in Table~4. Figure~3 shows the model fits in green, avoiding the CaT lines. Many low-S/N spectra still had sky-line contamination past 8800\AA -we found more consistency in the model fits when avoiding this region.

\subsubsection{Comparing Metallicity Scales}

Our CaT metallicities, calibrated with the \citet{dias20} method, are tied to the
\citet{cole04} CaT-relation and therefore effectively lie on the modern \citet{carretta97} / \citet{kraft03} GC scale, which
is closely consistent (to within $\sim 0.05$-$0.1$ dex at
${\rm [Fe/H]} \simeq -0.8$) with the \citet{carretta09} / \citet{harris10}
compilation.  The P23 metallicity for NGC~6569 adopts the
\citet{valenti11} high-resolution near-IR scale, which is likewise matched
to the \citet{carretta09} / \citet{harris10} values.  The \citet{crociati23} CaT measurements are placed on the \citet{dias16} CaT scale via the
\citet{husser20} calibration, while the J18 metallicities
are on the \citet{carretta09} / \citet{harris10} scale through the
\citet{mauro14} CaT calibration and their homogeneous high-resolution
analysis. The J18 and P23 / \citet{crociati23} datasets overlap via the FLAMES spectra, at least. 

Figures~5a \& 5b compare the CaT-based calibrations of J18 and \citet{crociati23}, along with the \citet{dias20} scale and our \spc\ fits in Figure~5c. Out of 303 AAT spectra, 281 stars had successful \spc\ fits, but only 58 fall within the RV range of NGC~6569 and were suitable for comparison. Even before applying metallicity filters, a Kolmogorov-Smirnov (K-S) test showed that the RV-selected samples (AAT and J18) are statistically indistinguishable (t-statistic = $-1.32$, p-value = 0.1880). The mean [Fe/H] from our recalculation of the J18 sample's \fedp\ was $-0.83 \pm 0.15$~dex (98 stars, only slightly lower than the J18 value of -0.87 dex, using a $K$-band luminosity correction), while our mean for the 58 AAT stars was \fedp\ $= -0.90 \pm 0.40$~dex (t = 1.27, p = 0.2076), suggesting no significant difference. Our values also agree with those from \citet{crociati23}, who used a V-band HB luminosity correction. We conclude that all three CaT calibrations are on a consistent metallicity scale.

We further compared \spc\ [Fe/H] values (\fesp ) with CaT-based estimates (\fedp) for the RV-selected sample (Figure~5c). Some of the 58 extra-tidal candidates may be closer to us than NGC~6569, rendering the ($i-i_{HB}$) luminosity correction uncertain. J18 noted that CaT-based [Fe/H] is accurate to within $\pm 0.3$~dex per star. We find that \fedp\ matches \fesp\ within this range for stars with $-1.1 < $ \fedp\ $< -0.5$~dex (highlighted as orange-filled green circles; red dotted line in Figure~5c). The absolute difference between methods is reported in Table~4 as $\Delta [\mathrm{Fe/H}] =[\mathrm{Fe/H}]_{SP} - [\mathrm{Fe/H}]_{DP}|$. We also note that reliable \textit{Gaia} parallaxes exist only for stars with Plx uncertainties $< 20$\%  
(8 stars); other distances are marginal or unreliable. 

A mild offset with $\mathrm{[Fe/H]}_{\rm CaT}<\mathrm{[Fe/H]}_{\rm SP}$ is expected for some giants because the CaT calibration hinges on Ca-line EWs and a photometric luminosity correction $(i-i_{\rm HB})$; small errors in distance/reddening or in $\log g$ (e.g., AGB/RGB mix), and any mismatch in $[\mathrm{Ca/Fe}]$ relative to the calibration set, systematically reduce the CaT-inferred iron abundance, whereas SP\_Ace fits many Fe and $\alpha$ lines simultaneously.

Figure~6 shows the CMD selection for targets that meet the RV-range and the [Fe/H]-limits; this is the dereddened BDBS CMD in $i\; vs. \; (u-i)$ for the inner-cluster members (gray filled circles); the BDBS detected 98/100 stars from the J18 sample are shown as small green pluses; overplotted are the objects A-D are shown (Table~4). The grades are assigned as follows: A: Large green pluses with a black border have $-0.5 >$ \fesp $> -1.1$~dex, and  $\mathrm{[ \alpha/Fe]} > +0.2$~dex.;
Grade B: Smaller blue pluses have $-0.5 > \text{[Fe/H]} > -1.1$~dex and $\mathrm{[ \alpha/Fe]} > +0.1$~dex. Grade C: small red pluses have $-0.5 > \mathrm{[Fe/H]} > -1.1$~dex and $-0.1 < \mathrm{[ \alpha/Fe]} < +0.1$~dex, or objects with only $\mathrm{[Fe/H]}_{DP}$ in the correct range. Grade D: stars with $\Delta [\mathrm{Fe/H}] > 0.30$~dex and $-0.5 > \mathrm{[Fe/H]_{SP}} > -1.1$~dex.

Throughout this paper, we use the 98/100 J18 RGB stars (with recovered PMs) for comparison. This sample is consistent with the expected $N_{\rm RGB}\!\sim\!120$-$150$ for a $12$~Gyr, $[\mathrm{Fe/H}]\!\approx\!-0.85$~dex cluster of present mass $M_{\rm cl}\!\approx\!2.3\times10^{5}M_\odot$, once we account for (i) crowding/incompleteness in the inner $\lesssim2'$ (Figure~1 and discussion of PM selection), (ii) our magnitude cuts and PM box, and (iii) the $\sim$35\% field-interloper fraction in the PM-selected set (\tt{ASteCA}'s \rm CI$=0.35$). Together, these effects naturally lower the recovered PM-vetted giant count from the theoretical expectation to the observed $N\!=\!98$.

The CaT calibrations used here  \citep{dias20, crociati23} include formal uncertainties, but their accuracy degrades when applied to stars that differ significantly from the calibration sample in terms of $\alpha$-enhancement, gravity, microturbulence, effective temperature, or continuum placement. This often biases [Fe/H] values for bulge/disk RGB stars toward artificially low metallicities. In contrast, full spectral synthesis methods such as \spc\ model line blending, solve self-consistently for atmospheric parameters, and fit dozens of Fe lines simultaneously (given $S/N > 20$), yielding more robust and accurate [Fe/H]-estimates. The \spc\ values are calibrated onto a solar-relative scale anchored to the \citet{grevesse98} solar composition and the high-resolution abundances of \textit{Gaia} benchmark stars \citep{ramirez11, jofre14}, rather than to any specific globular-cluster metallicity compilation.

\begin{figure*}[ht!]
\centering
\includegraphics[trim={0 0 0 0},clip,width=.95\textwidth]{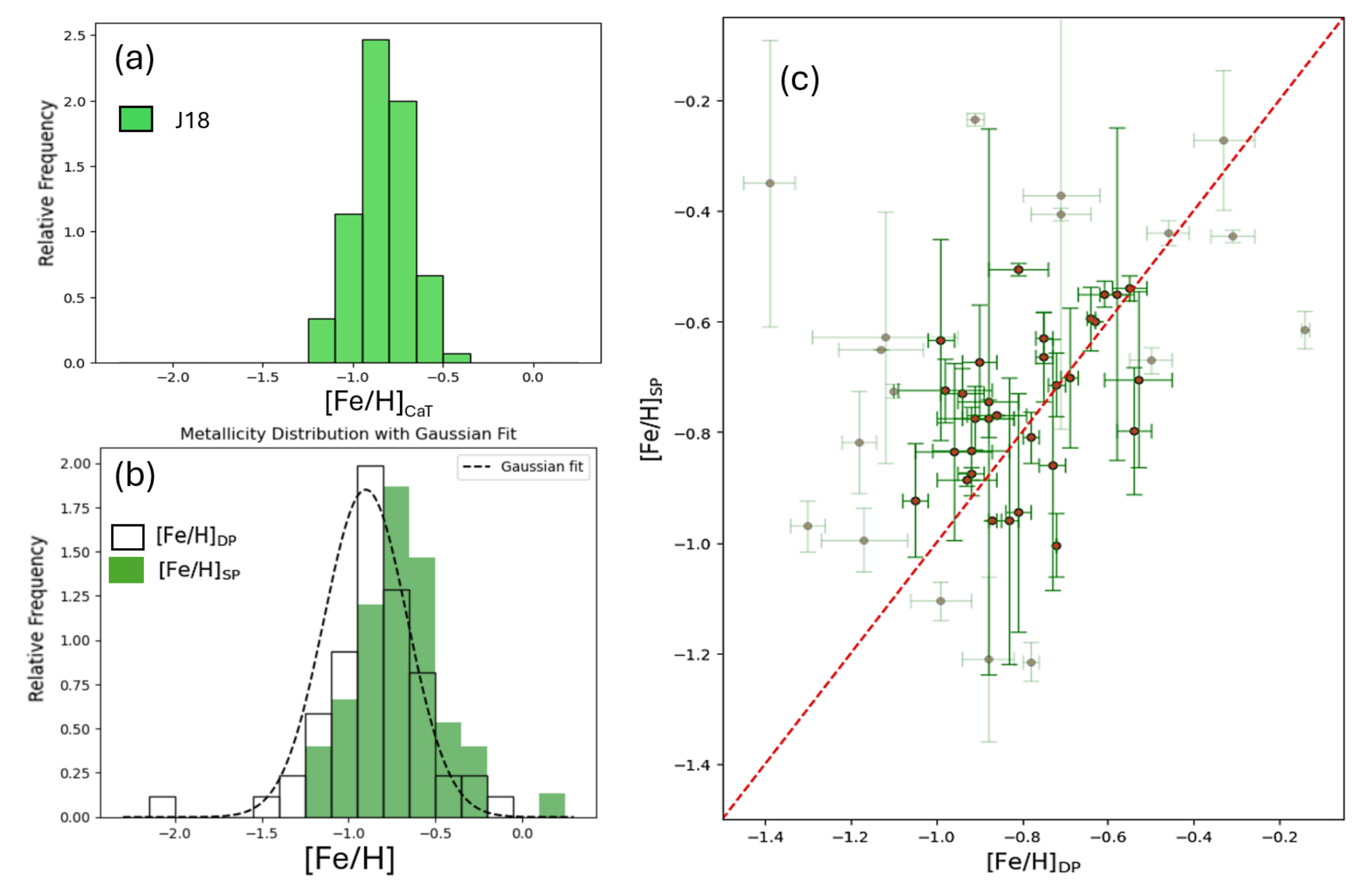}
\caption{Comparisons between the calibration methods of obtaining [Fe/H] from the CaT lines \citep[J18]{dias20} with the \spc\ code results. \textbf{(a)} A histogram of the J18 sample of  98/100 stars is shown in green, which were matched with BDBS photometry. For these 98 stars, using the luminosity correction in the i-band, we find:
$\textrm{[Fe/H]}_{CaT}= -0.84\pm 0.17$~dex. \textbf{(b)} The 58 stars in the RV-selected sample ($-25>RV>-65$~\kms ) are shown as the black-line histogram (no fill), calibrated using \citet{dias20} method. For all objects: $\textrm{[Fe/H]}_{DP} = -0.91\pm 0.40$~dex. Targets with $3500<T(K) <7500$ were fit with the \spc\ code: $\textrm{[Fe/H]}_{SP} = -0.78\pm 0.20$~dex. The dotted line is the Gaussian fit to the P23 data, presented in \citet{crociati23}: $\textrm{[Fe/H]}_{CaT} = -0.90\pm 0.20$~dex. \textbf{(c)} Comparing the two fits for the overlap between the samples (all stars are faint green and orange filled circles). The red line is the 1:1 fit of the CaT-method and the \spc\ code, which fits the stars with both  \fedp\ and \fesp\  are limited to values between -0.5 and -1.1~dex (40 objects; brighter green and orange points).
 } 
\end{figure*}

\begin{figure*}[t]
\centering
\includegraphics[width=\textwidth]{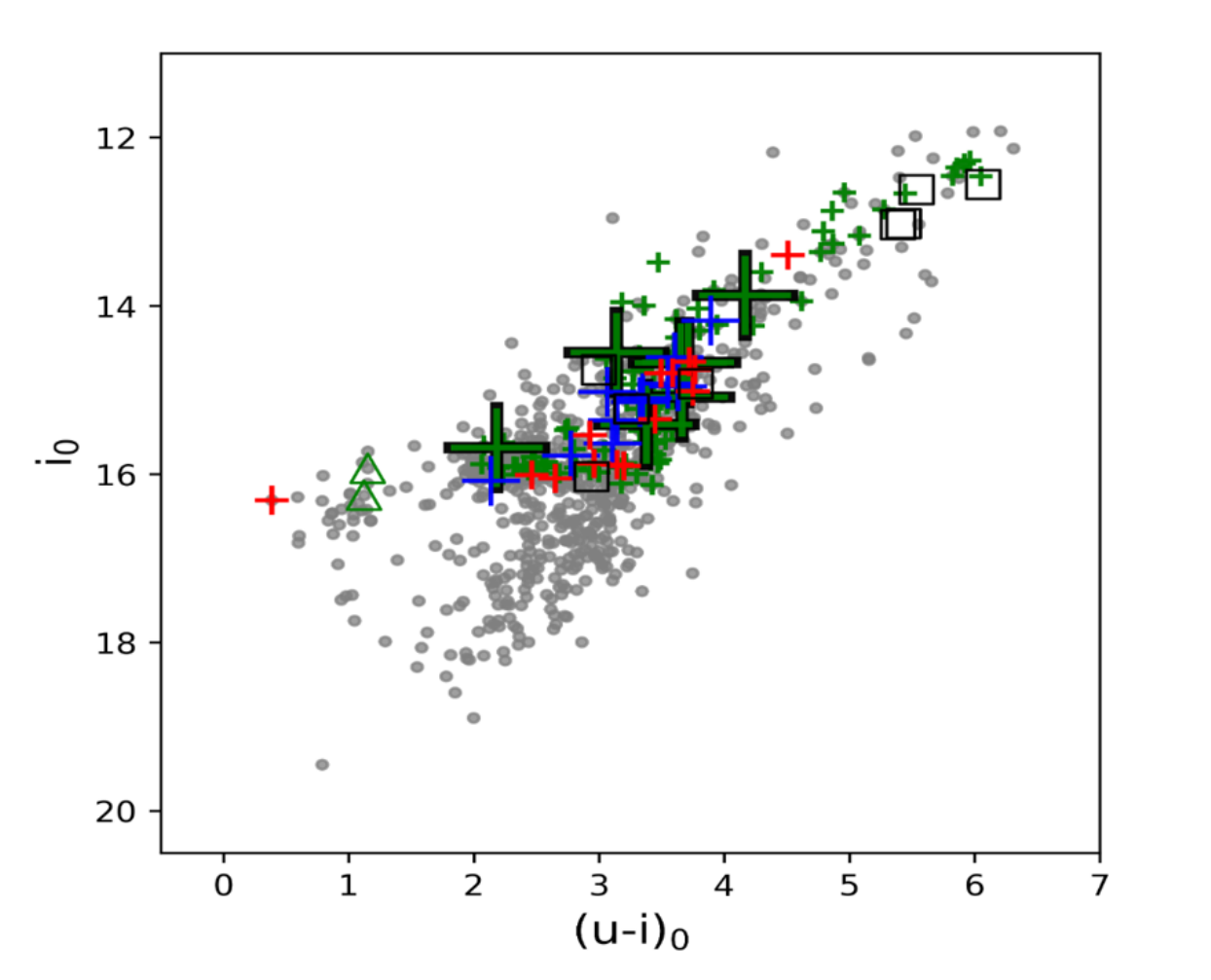}
\caption{Candidate selection for tidal debris around NGC~6569: The deredded i vs. (u-i) BDBS CMD is shown for objects ASteCA labeled as cluster members (gray filled circles). The BDBS-detected 98/100 stars from the J18 sample are shown as small green pluses. The objects A-D are shown (Table~4). Grades
A: Large green pluses with a black border;
Grade B: Smaller blue pluses;
Grade C: Small red pluses;
Grade D: Open black squares. }
\label{fig:selection}
\end{figure*}

\subsection{Extra-Tidal Candidates}

In our AAT sample, 58 stars ($19\pm 4$\%) fall within the RV range $-65 < RV < -25$~\kms , making them initial candidates for stars recently escaped from the cluster. Of these, 40 meet the [Fe/H] range $-1.1 < \text{[Fe/H]} < -0.5$ dex (Figure~5c) using \fesp, and 50 are within this range using the CaT method only (\fedp). Figures~6 \& 7 show the selection of extra-tidal candidates by RV, \fedp , \fesp , and \afe . We were able to measure \fesp\ and \afe\ for 281 stars (6200-3600K), 58 of which were in the RV-range for NGC~6569's recently escaped population. Figure~6 shows the objects selected as extra-tidal candidates, based on the criteria displayed in Table~4. 

\begin{figure*}[ht!]
\centering
\includegraphics[width=1.0\textwidth, height=0.425\textheight]{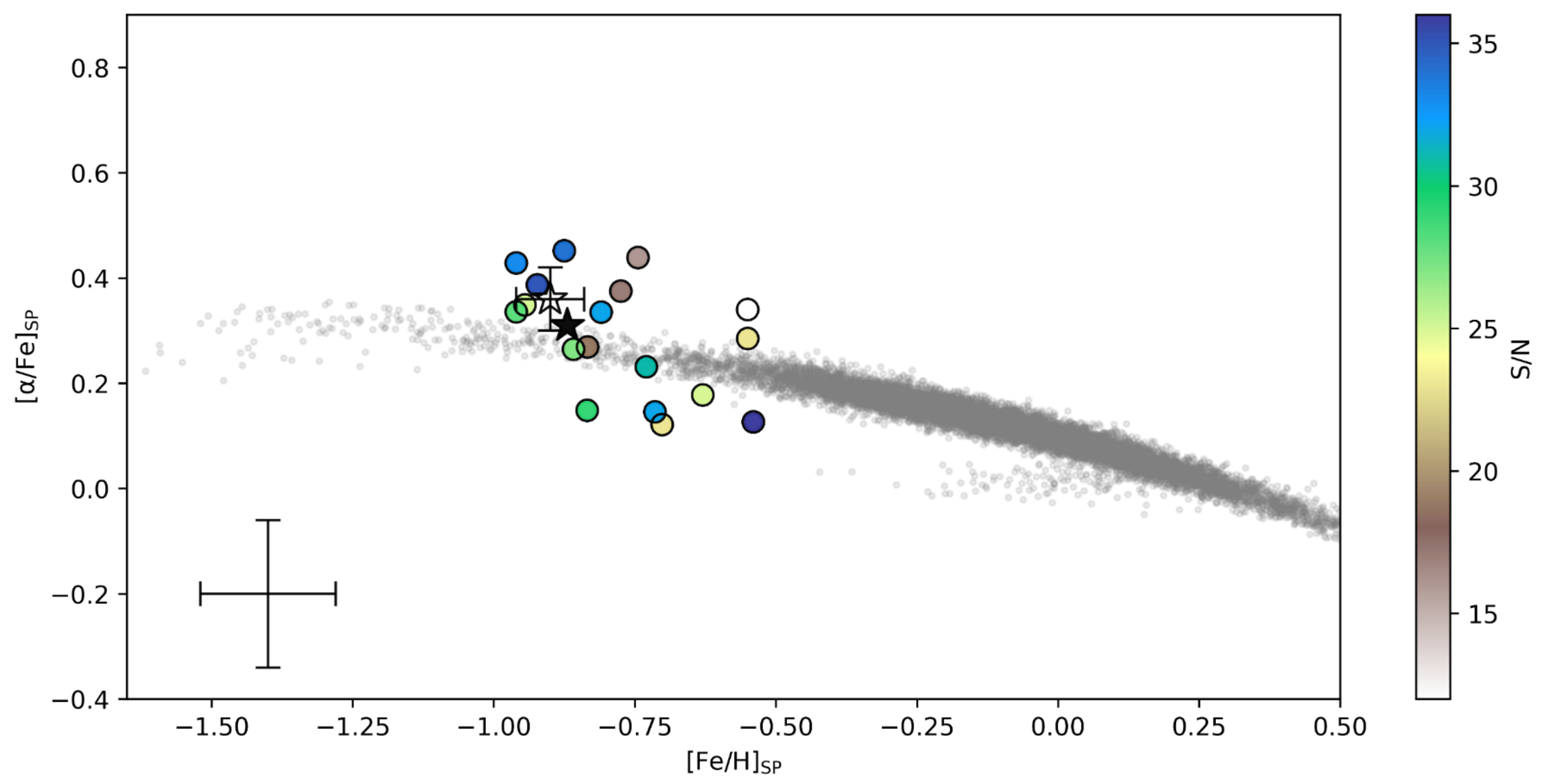}
\caption{The \spc\ code-derived [$\alpha$/Fe] versus [Fe/H] for the best extra-tidal candidates selected in Figure~6, grades A+B (19 objects). The black-filled star marks NGC~6569's mean values from high-resolution samples (see J18). The open star with error bars is the average of the \citet{barrera25} sample. Within uncertainties, candidates follow a single locus consistent with NGC~6569; we find no compelling evidence for intrinsic spreads in [Fe/H] or [$\alpha$/Fe]. A few high-[$\alpha$/Fe] points occur at low S/N and are treated as upper limits. \rev{The gray dots are the \bes\ Galactic population models generated for our AAT FOV and PM-limits.}}
\label{fig:alpha_fe}
\end{figure*}

\begin{figure*}[ht!]
\centering
\includegraphics[width=1.0\textwidth, height=0.33\textheight]{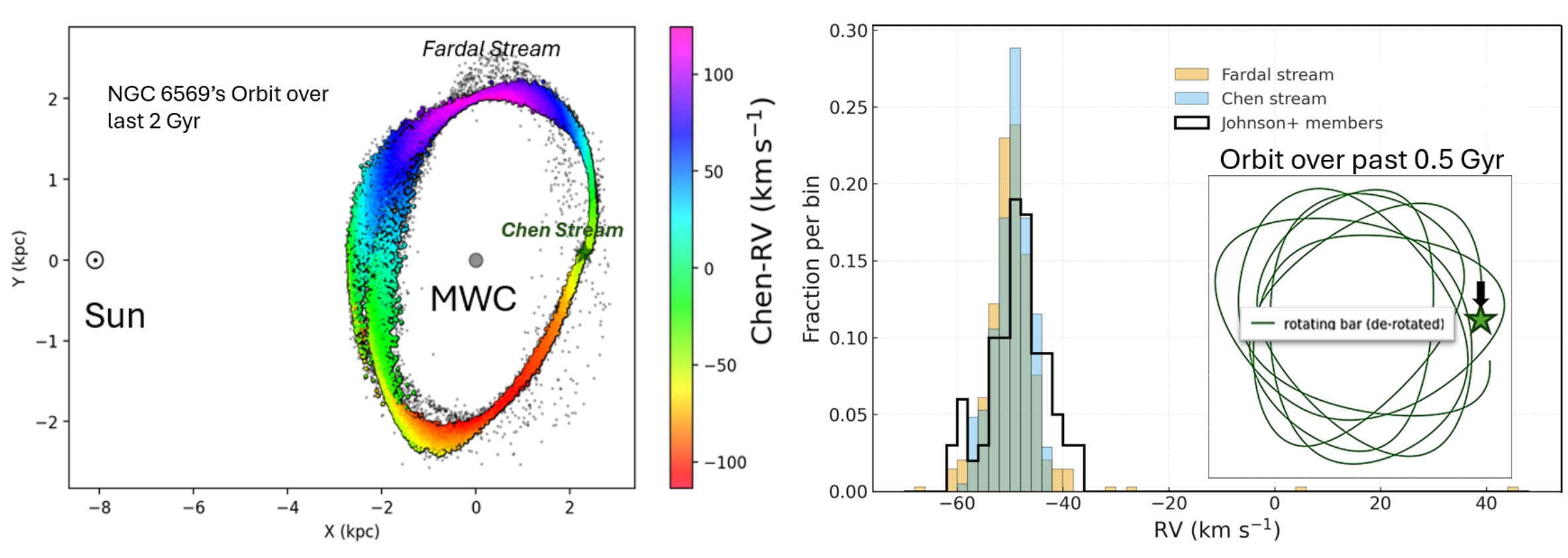}
\caption{\bf Left Panel: \rm The \textsc{Gala}-generated mock-tidal streams are shown in the Galactocentric frame (MW center is shown as the gray filled circle), where the black-bordered green-filled star is the position of NGC~6569 at the present epoch coordinates X,Y,Z, are (2.303, 0.087, -1.229)~kpc. Our position is shown as the sun symbol. \textsc{Gala} was set to have star particles ejected every 0.5~Myr for 2.0~Gyr. The \citet{chen25} mock stream star particles are color-coded by the heliocentric RV (right color bars), and the \citet{fardal15} star particles are shown as the gray dots. \bf Right Panel: \rm The relative number histograms are shown for the J18-observed CaT-sample stars (heavy black outline), the Chen mock-stream star particles (aqua shading), and the Fardal mock stream (sand-colored shading), released in the last 0.5~Gyr, having our AAT-selection of PM, and found in our FOV.  The inset shows the orbit for the last 0.5~Gyr, with the motion shown by the black arrow above the star cluster (black-bordered green star). The cluster was last at pericenter $\sim 48.5\,Myr$ ago.
 }
\end{figure*}

 We find that only stars with $S/N > 35$ or $\alpha$-enhancement
 \afe\ $> +0.20$~dex yield reliable abundance measurements other than Fe; lower-quality spectra provide only upper limits for \afe. At lower signal-to-noise ratios, the [$\alpha$/Fe]-abundances derived using \spc\ become increasingly unreliable. In this regime, the absorption lines of $\alpha$-elements (e.g., Mg, Si, Ca, Ti) are often indistinguishable from the noise, causing the spectral synthesis algorithm to misidentify features or default to prior assumptions embedded in the model library. This tends to \textit{overestimate} $\alpha$-enhancement, particularly when the continuum is poorly constrained. Therefore, [$\alpha$/Fe] measurements for stars with $S/N < 35$ should be treated as \textit{upper limits}. This bias is apparent in Figure~7, where a number of faint stars with low S/N exhibit artificially elevated [$\alpha$/Fe] values, inconsistent with the trends seen in higher-quality spectra of cluster members.

Among the Grade~A$+$B candidates ($N=19$; shown as S/N-color-coded filled circles in Figure~7), the observed dispersions are
$\sigma_{\rm obs}([\mathrm{Fe/H}]_{\rm SP})=0.14$~dex and
$\sigma_{\rm obs}([\alpha/\mathrm{Fe}]_{\rm SP})=0.14$~dex, while the RMS measurement uncertainties are
$\sigma_{\rm meas}([\mathrm{Fe/H}]_{\rm SP})=0.17$~dex and
$\sigma_{\rm meas}([\alpha/\mathrm{Fe}]_{\rm SP})=0.25$~dex.
Subtracting in quadrature,
$\sigma_{\rm int}=\sqrt{\max\!\left(0,\sigma_{\rm obs}^2-\sigma_{\rm meas}^2\right)}$,
yields $\sigma_{\rm int}\approx 0$~dex in both dimensions (i.e., no detectable intrinsic spread at the level of our current measurement uncertainties).
To quote a conservative upper limit on any intrinsic metallicity scatter, we take the one-sided 95\% upper confidence bound on the true dispersion for $N=19$,
\begin{equation}
\sigma_{95}
=\sigma_{\rm obs}\,\sqrt{\frac{N-1}{\chi^2_{0.05,\,N-1}}}
=1.38\,\sigma_{\rm obs}
=0.19~{\rm dex},
\end{equation}
and propagate the measurement uncertainty to obtain
$\sigma_{\rm int}([\mathrm{Fe/H}]_{\rm SP})<\sqrt{\max\!\left(0,\sigma_{95}^2-\sigma_{\rm meas}^2\right)}\simeq0.074$~dex.
For $[\alpha/\mathrm{Fe}]_{\rm SP}$, the typical uncertainties ($\sim0.25$~dex) dominate, so we do not place a meaningful upper limit below the measurement floor (conservatively, $\sigma_{\rm int}([\alpha/\mathrm{Fe}]_{\rm SP})\lesssim0.25$~dex).

Comparing with the expected field-star contamination, the 7,562 model stars contained in the \bes\ sample (gray dots in Figure~7) for our FOV were selected using the same PM criteria as the spectroscopic target objects, with BDBS {\it i}-band magnitudes above the approximate brightness limit of 17.5 magnitudes. Selecting for the heliocentric radial velocity (HRV) limits of $-25$ to $-65 $~\kms , leaves 1,660 objects (22\%), but there are only 539 model stars with $-1.1<[Fe/H]<0.5$ dex, and   \afe $>+0.2$ dex, which is 7\%. We found 58/303 of our AAT spectra in the same HRV range, which is 19\%, and 41 (14\%) stars were classified as having passing grades of ABCD from Table 4's Notes. 

To numerically assess the likelihood that individual stars belong to the globular cluster NGC~6569 or its associated tidal debris, we computed a set of membership probabilities based on a weighted scoring model incorporating multiple stellar parameters. The scoring system prioritized stars with RVs in the ABCD-range $-30$ to $-65$~kms$^{-1}$ (comparing well with J18), consistent with theoretical expectations for cluster and tidal tail members. Additional weights were assigned based on metallicity indicators, including \fesp\ and \fedp , as well as $ \alpha$-element enhancement, signal-to-noise ratio (S/N), and the difference between the \spc\ and CaT Fe-estimates, where available. Stars with high S/N ($>35$) and temperatures in the range $3500K < T < 7500K$ were evaluated using \fesp , while stars outside this range, where reliable stellar parameters could not be derived, were evaluated using \fedp\ alone. Membership scores were normalized to a scale of $0-1$ and interpreted as relative membership probabilities. This approach allowed us to incorporate both kinematic and chemical information, providing a multidimensional assessment of the likelihood of cluster membership. While these probabilities help identify the most likely members, we emphasize that they are model-dependent and should be interpreted in the context of the estimated contamination index (Eqn.[1] and \S2.2: CI = 0.35), which quantifies the expected field star contamination within the selection region. The ASteCA contamination index of $CI = 0.35$ implies that approximately a third of the stars within the selection region may be field interlopers. While our membership probability model accounts for kinematic and chemical constraints, this CI highlights the inherent risk of contamination in any cluster-centric selection. Therefore, membership probabilities should be interpreted relative to this baseline contamination rate, and caution should be applied when assigning absolute certainty to individual stars based solely on computed probabilities. 

In Figures~6 \& 7, there are 7 stars graded A ($-0.5 >$\fesp$ > -1.1$ dex, and $\mathrm{[ \alpha/Fe]} > +0.2$ dex), shown as large bright green plus symbols, also having $\Delta [\mathrm{Fe/H}] = |\mathrm{[Fe/H]}_{SP} - \mathrm{[Fe/H]}_{DP}| < 0.15$ dex. There are 12 stars graded B ($-0.5 > \mathrm{[Fe/H]} > -1.1$ dex and $\mathrm{[ \alpha/Fe]} > +0.1$ dex) depicted as smaller blue plus signs. Stars graded C include those with $-0.5 > \mathrm{[Fe/H]} > -1.1$ dex and $-0.1 < \mathrm{[ \alpha/Fe]} < +0.1$ dex, or objects with only $\mathrm{[Fe/H]}_{DP}$ in the correct range, and are shown as small red plus-signs. Objects with $\Delta [\mathrm{Fe/H}] > 0.30$ dex are represented by small black square borders (grade D), where \citet{cole04}  point out that the presence of systematic errors in CaT measurements can be exacerbated in environments typical of RGB stars near the RGB tip, where the stellar parameters are often less stable, or they could be closer to us than NGC~6569. J18's sample are shown as small green pluses in Figure~6. The extra-tidal candidates are consistent with the bound cluster population of J18. We compare the derived parameters for the selection of AAT-stars (graded ABCD) with the J18 and P23 samples in Table 5, including the ESO spectra that were available publicly. 

To investigate the D-class objects further, we took a typical RGB stellar model with $\textrm{[Fe/H]} \approx -0.8$ dex at a distance of $\sim 10.5$~kpc and adjusted the distance to 6~kpc. An erroneous luminosity correction to the assumed HB position for the center of NGC~6569 would return a value \fedp\ $\leq -1.0 $~dex. Such stars would appear brighter than we expect, which may be the source of the double HB reported by Mauro et al. (2012), in the {\it K}-band, and suggested by the dereddened BDBS CMDs shown in Figures~2 \& 6. However, if we assume all the double HB stars are members, the magnitude difference is wavelength dependent (J18), which is more indicative of a composition difference (e.g., He-abundance). We note that the upper-RGB stars with $\Delta [Fe/H] > 0.3$~dex may be variable, causing the discrepancy. Future work will involve N-body simulations, which will be able to estimate the stellar masses of escaped stars, rather than just the {\it massless} mock stream particles.

\section{Theoretical Modeling and Comparison with Observations}

In this work, we make complementary use of the \textsc{agama} and \textsc{gala}
codes.  \textsc{agama} \citep{vasiliev19} is a C++/Python framework for
action-based galaxy modelling and self-consistent, multi-component Galactic
potentials; we use it to compute NGC~6569's orbit, Jacobi radii, and escape
velocities in a barred inner-Galaxy potential.  \textsc{gala}
\citep{pricewhelan17} is a Python dynamics toolkit optimized for orbit integration
and mock stream experiments; we adopt it to integrate orbits in the chosen
potential and to generate and visualize tidal-debris models that can be
directly compared with our chemo-dynamical sample. We explicitly compare the
orbits of NGC~6569 computed with \textsc{agama} and \textsc{gala} in the Appendix.  The resulting pericenter and apocenter distances,
vertical excursions, orbital period, and time-dependent Jacobi radii
agree to within the small numerical differences expected from the
different integrators and internal implementations, and are well below
our observational and modeling systematics.

We compared the properties of our candidate extra-tidal stars with 2 mock tidal-stream models \citep{fardal15, chen25} generated using the \textsc{Gala} Python package \citep{pricewhelan17, pricewhelan22}. 
The \citet{fardal15} method (emphasizing phase-space sampling that preserves broad stream statistics) and the \citet{chen25} method (yielding sharper morphology and kinematics). In our case (Figure 8), the Chen realization produces physically narrower tidal tails, with a smaller range in RV. Details are provided in the Appendix.

Our fiducial MW model combines: a spherical Navarro, Frenk \& White (NFW) halo \citep{navarro96,navarro97} with circular speed $V_c=220$~\kms\ at the scale radius $r_s=15~\mathrm{kpc}$; a Miyamoto-Nagai disk \citep{miyamoto75}; and a triaxial bar following \citet{long92}, tilted by $25^\circ$ from the $x$-axis, with mass $M_{\rm bar}=(1/6)\,M_{\rm disk}$, long-axis scale length $4~\mathrm{kpc}$, and pattern speed $\boldsymbol{\Omega}=(0,0,42)\,\mathrm{km\,s^{-1}\,kpc^{-1}}$ \citep{sanders19,bovy19}.

NGC~6569 was modeled as a Plummer potential \citep{plummer1911} with its current mass $M=2.3\times10^{5}\,M_\odot$ \citep[Table 1;][]{vasiliev21}. Varying the global MW potential shifts debris only at separations of several degrees from the cluster; Within a few degrees, the tail \emph{shape} is governed mainly by the assumed cluster mass, and a smaller cluster mass shortens the leading tail. Changing the cluster mass alters the width/length of the tail, but not the debris's \emph{direction}. \textsc{Gala} orbits for NGC~6569 remain confined to the bulge (Figure~8-left), consistent with previous work (for example, \citealt{munoz17}; J18; P23).

We also tested the models with P23's parameters, adopting the smaller cluster mass and closer distance in the
\textsc{gala} spray models, the resulting leading tail is noticeably shorter, but the ejected debris directions are consistent with the values we chose.
This behavior is expected: a lower $M_{\rm cl}$ reduces the Jacobi radius
$r_J \propto (M_{\rm cl}/M_{\rm enc})^{1/3}$ and the internal velocity
dispersion of the cluster, so stars escape with smaller energy and action
offsets and therefore drift more slowly along the orbit. 

In the Appendix, we quantify the fiducial MW model's effect on orbital evolution over the last few Gyr: Figure~8 shows an orbital integration for both 2~Gyrs and the last 0.5~Gyr. The cluster is placed at its cataloged location; the code then takes the cluster backward in time with its current mass, then sends it forward in time, spitting out mock-stream particles, shown with the heliocentric RVs predicted for the escaping \emph{massless star particles}, which then follow their own orbits once away from NGC~6569's influence. The right-panel inset shows the Galactocentric orbit: the current cluster position (green star) and forward motion (black arrow) are indicated. During the forward leg, particles are ``sprayed" every $0.5$~Myr. This cadence is justified by the dynamical crossing time,
\begin{equation}
t_{\rm cr}=\frac{r_h}{\sigma}\approx\frac{3.5~\mathrm{pc}}{7.5~\mathrm{km\,s^{-1}}}\approx0.42~\mathrm{Myr},
\end{equation}
using $r_h\simeq3.36\pm0.15$~pc and $\sigma\simeq7-8$~\kms~(Table~1; see also \citealt{valenti11}, J18, P23, and the Baumgardt \& Hilker catalog), following standard procedures in \citet{binney08}.

\textsc{Gala} predicts that PM-selected extra-tidal stars in our field of view (FOV) should largely share NGC~6569’s RV-range, and (Figure~8, right histograms). Where the observed distance is $10.53\pm0.26$~kpc (Table~1). Using our reasonable spray rates, $>95$\% of particles in our FOV escaped within the last $50$~Myr. The estimated orbital period is complex due to the rotating bar (see Figure A.1 and Table A.1). Longer runs showed that earlier escapees using the Fardal particle-spray can appear in our PM window at smaller heliocentric distances ($\sim6-8$~kpc) with $-220\lesssim \mathrm{RV}\lesssim -100$~\kms, i.e., on the near side of the Galactic center. NGC~6569 may thus lose stars along much of its bulge orbit, but quantifying this requires full N-body modeling for the mass-loss rate.

The RV-histograms shown in the right panel of Figure~8 indicate the J18 accepted cluster members have a greater velocity dispersion than the recently published \citet{chen25}-generated mock stream for our FOV limits. Again, we compare our AAT sample with the J18 sample to determine the expected properties of escaping stars. To find the bound/unbound status of our AAT dataset, we evaluated the probability of being unbound, the P-value ($\equiv P_{\rm unbound}$) in Table~4 (Appendix Eqn.[A.7]), for both the J18 members and our AAT extra-tidal candidates. This method uses a dispersion-aware (taken from P23) escape criterion anchored to the cluster’s Jacobi surface. For each star, we form the cluster-rest-frame 3-D speed from \textit{Gaia} DR3 proper motions and line-of-sight velocities at a fixed cluster distance of $D=10.53$ kpc, after subtracting the systemic motion $(\mu_{\alpha*},\mu_\delta)=(-4.142,-7.333)\,\mathrm{mas\,yr^{-1}}$ and $v_{\rm sys}=-49.8\,\mathrm{km\,s^{-1}}$. We compare $|{\bf v}_{\rm rel}|$ to an effective escape-speed curve from a Plummer potential (scale $a=r_h/1.305$) truncated at the \textsc{AGAMA}-derived Jacobi radius $r_J=14\farcm24$ (43.6 pc, Appendix Eqns.[2] \& [3]), and inflate the threshold by the cluster’s inner-plateau dispersion, adopting $\sigma_{\rm los}=6.5\,\mathrm{km\,s^{-1}}$ so that $v_{\rm thr}^2=v_{\rm esc,eff}^2+(n\,\sigma_{3\rm D})^2$ with $\sigma_{3\rm D}=\sqrt{3}\,\sigma_{\rm los}$ and $n=2.5$. Measurement errors are propagated via Monte-Carlo draws ($N\!\sim\!10^3$) to compute $P_{\rm unbound}\equiv\Pr(|{\bf v}_{\rm rel}|>v_{\rm thr})$, and we label stars as \emph{bound} ($P_{\rm unbound}\le0.2$), \emph{unbound} ($P_{\rm unbound}\ge0.8$), or \emph{borderline} (otherwise). In Appendix A (Table A.4), the 98 stars in the J18 sample divide into 83 bound stars, with 4 borderline cases, and 11 unbound stars. Our “dispersion-aware” boundness is the velocity form of the Jacobi criterion:
$E_J=\Phi_{\rm eff}+\tfrac12 v^2 \ge E_{J,{\rm crit}}$ $\Leftrightarrow$ 
$v\ge v_{\rm esc,eff}(R)$, with our stated assumptions.

In the axisymmetric \textsc{gala} potential, we obtain
$r_J \approx 11\arcmin$-$18.3\arcmin$, whereas the full barred
\textsc{agama} model yields a smaller Jacobi radius at pericenter
($r_J \sim 8\arcmin$-$10\arcmin$) and a larger one at apocenter
($r_J \sim 18$-$22\arcmin$), corresponding to an effective ``breathing''
range of $\sim 11\arcmin$-$19\arcmin$ around a mean value of $\sim 14\arcmin$.
Thus, P23's determination of the ``tidal" radius is consistent with the \textsc{agama} model's inner limits.

Applying the same procedure to our AAT RV- and metallicity-selected sample enables a direct, internally consistent comparison with the J18 members (and thus a subset of the P23 sample: Table A.5) used for all previous comparisons. Of the Grade-A objects, all have a similar range in RV, PM, and chemical composition as the J18 sample, and some of them appear still bound to NGC~6569, as is shown in Figure~9. Indeed, most of the AAT sample inside the Roche surface (dotted lines between L1 \& L2) appears still bound to the GC. Of the other classes, BCD, a smaller percentage is unbound, but that decreases with the radial distance. The RRL subsample was only included for completeness.

\begin{deluxetable*}{clllllllllllllllllll}
\tabletypesize{\tiny}
\tablewidth{0pt} 
\tablecaption{Extra-Tidal Candidates \& Non-Members}
\tablehead{\\
\colhead{No.} & \colhead{[M/H]} & \colhead{$\sigma$} &\colhead{\fesp} & \colhead{$\sigma$} &\colhead{ \afe} & \colhead{$\sigma$} &
 \colhead{T} & \colhead{$\sigma$} &\colhead{Log g} & \colhead{$\sigma$} &\colhead{RV} &  \colhead{$\sigma$} &\colhead{S/N} & \colhead{\fedp $^*$} &  \colhead{$\sigma$} &\colhead{$\Delta [Fe/H]$} &
 \colhead{Class} & \colhead{Prob.}& P\\
 \colhead{(1)} & \colhead{(2)} & \colhead{(3)} & \colhead{(4)} & 
 \colhead{(5)} & \colhead{(6)}& \colhead{(7)} & \colhead{(8)}& \colhead{(9)} &\colhead{(10)}& \colhead{(11)} & \colhead{(12)}&\colhead{(13)} & \colhead{(14)}& \colhead{(15)} & \colhead{(16)}&
 \colhead{(17)} & \colhead{(18)} & \colhead{(19)} & \colhead{(20)} }
\startdata 
\\
84& -0.53& 0.03& -0.87& 0.01& 0.45& 0.01& 4420& 26& 1.88& 0.06& -61.01& 2.43& 34& -0.92& 0.03& 0.05& A& 1.00& 0.85\\
86& -0.32& 0.14& -0.55& 0.02& 0.28& 0.14& 4865& 37& 2.64& 0.33& -55.32& 2.45& 23& -0.61& 0.06& 0.06& A& 1.00& 0.93\\
145& -0.62& 0.09& -0.92& 0.10& 0.39& 0.08& 4928& 99& 2.76& 0.18& -38.34& 2.54& 35& -1.05& 0.03& 0.13& A& 1.00& 0.08\\
221& -0.70& 0.08& -0.96& 0.26& 0.43& 0.21& 4258& 50& 1.91& 0.76& -61.08& 2.38& 33& -0.83& 0.02& 0.13& A& 1.00& 0.58\\
237& -0.67& 0.08& -0.94& 0.22& 0.35& 0.21& 4576& 72& 2.85& 0.75& -41.11& 2.11& 25& -0.81& 0.03& 0.13& A& 1.00& 0.59\\
\multicolumn{19}{l}{\it Complete table available online, ranked by membership probability.}
\\
\enddata
\tablecomments{
$^*$Assuming that all stars are at the same distance of 10.53~kpc.\\
(1) AAT AAOmega fiber number.\\
(2) [M/H] from \spc\ code in dex.\\
(3) Uncertainty in [M/H] from \spc\ code in dex.\\
(4) [Fe/H] from \spc\ code in dex.\\
(5) Uncertainty in [Fe/H] from \spc\ code in dex.\\
(6) $[ \alpha/Fe]$ from \spc\ code in dex.\\
(7) Uncertainty in $[ \alpha/Fe]$ from \spc\ code in dex.\\
(8) Effective surface temperature from \spc\ code in K.\\
(9) Uncertainty in effective surface temperature from \spc\ code in K.\\
(10) $\log$ of surface gravity in dex from \spc\ code. \\
(11) Uncertainty in $\log$ of surface gravity in dex from \spc\ code. \\
(12) RV in \kms .\\
(13) Uncertainty in RV in \kms .\\
(14) S/N \\
(15) \fedp\ in dex,  found using \citet{dias20} method.\\
(16) Uncertainty in \fedp\ in dex,  found using \citet{dias20} method.\\
(17) $\left|[Fe/H]_{SP} -[Fe/H]_{DP}\right|$ \\
(18) Class:\\
Class A(7): $-25>RV>-65$~\kms , $-0.5 >$\fesp$ > -1.1$ dex, and  $\mathrm{[ \alpha/Fe]} > +0.2$ dex.\\
Class B(12): $-25>RV>-65$~\kms , $-0.5 > \mathrm{[Fe/H]} > -1.1$ dex and $\mathrm{[ \alpha/Fe]} > +0.1$ dex.\\
Class C(14): $-25>RV>-65$~\kms , $-0.5 > \mathrm{[Fe/H]} > -1.1$ dex and $-0.1 < \mathrm{[ \alpha/Fe]} < +0.1$ dex, or objects with only \fedp\ in the correct range.\\
Class D(8): $-25>RV>-65$~\kms , $-0.5 > \mathrm{[Fe/H]} > -1.1$ dex, but $\Delta [\mathrm{Fe/H}] > 0.30$ dex.\\
Class RR(4): $-25>RV>-65$~\kms , and known RRL stars.\\
Class F(258): Stars outside the accepted RV range for NGC 6569 or outside its metallicity range. \\
(19) Decimal relative probability of belonging to NGC~6569's center or tidal tails.\\
(20) Decimal relative probability of being unbound ($P_{unbound}$ from Eqn.[A.7]).
}
\end{deluxetable*}

\begin{figure*}[ht!]
\centering
\includegraphics[width=0.75\textwidth]{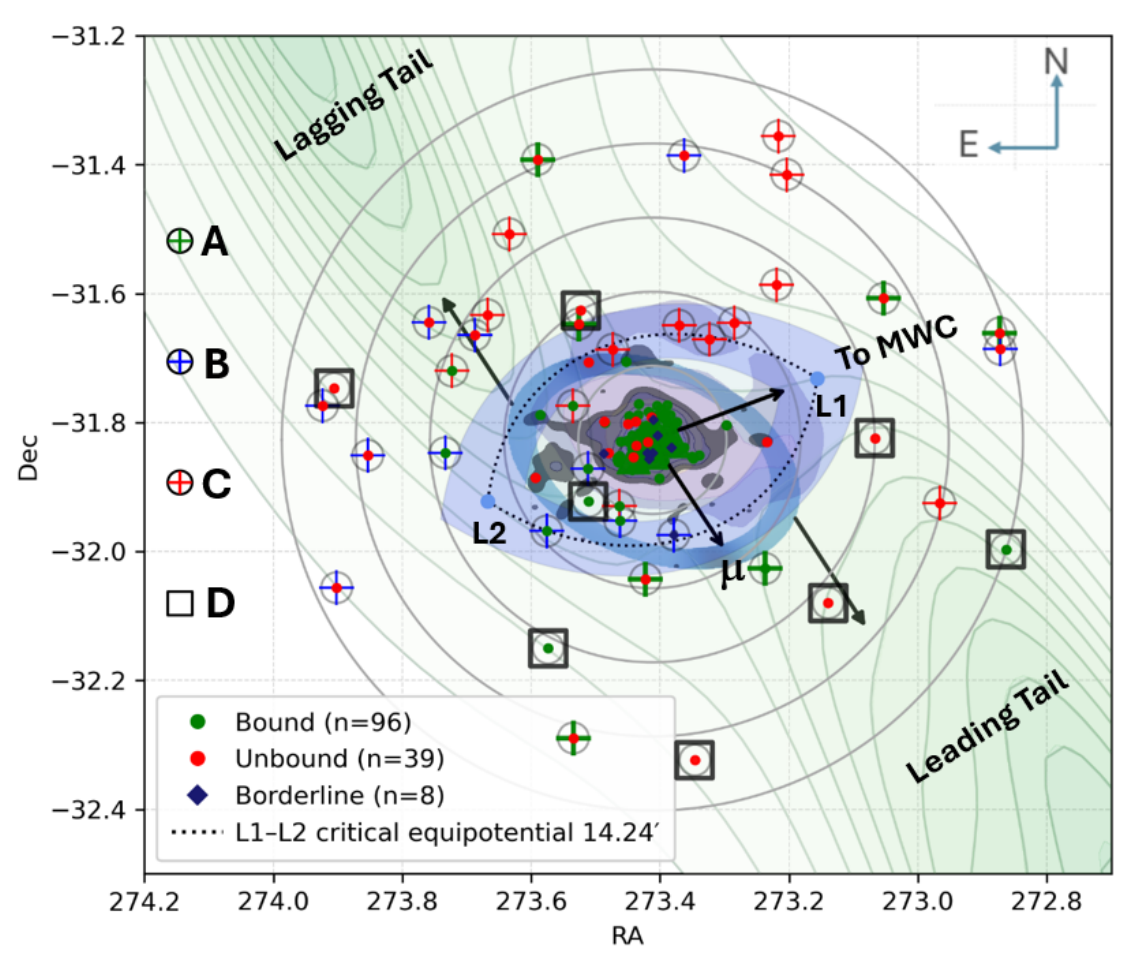}
\caption{
Observed AAT- and J18-datasets overlaid with \textsc{Gala}-generated mock-stream \citep{chen25}. Concentric gray circles mark $1$–$5\times r_t$ with $r_t=6\farcm9$ \citep{ortolani01,valenti11}. All AAT targets meeting the RV+metallicity cuts are shown as open circles, then filled with symbols identifying their properties. Symbol classes: 
A (7: large green pluses): $-1.1<\mathrm{[Fe/H]}<-0.5$ and $[\alpha/\mathrm{Fe}]>+0.2$; 
B (12: blue pluses): same RV, [Fe/H] with $[\alpha/\mathrm{Fe}]>+0.1$; 
C (13: red pluses): same RV, [Fe/H] with $+0.1>[\alpha/\mathrm{Fe}]>-0.1$ or CaT-only metallicities in range; 
D (8: open black squares): RV and [Fe/H] in range but possibly foreground ($d<10.5$~kpc). The J18 and AAT spectroscopic targets, in and around NGC 6569, are classified as energetically (see Appendix) \textit{bound} (green circles; $n=96$), \textit{unbound} (red circles; $n=39$), and \textit{borderline} (blue diamonds; $n=8$). The dotted curve marks the reference critical equipotential surface at $14\farcm24$, which is the Jacobi (tidal) surface or the Jacobi contour (at $E_J=E_{crit}$). Black arrows show the directions toward the Galactic Center (L1/L2 axis) and of the cluster proper motion $\boldsymbol{\mu}$. The Chen (rotating–bar, star-particle density) tidal debris is shown as green contours and shading (RV+PM window), the P23 spectroscopic sample (gray shading), and the \textit{Gaia} HB/RC selection (purple). The light blue shading surrounding the black dotted line indicates the L1/L2 Jacobi boundary between $11'$ and $18'$ over the whole orbit. The darker, blue-shaded, ellipse band is the most probable escape zone for stars over the whole orbit$-$stars would peel away from the critical equipotential near its flanks, not along the L1/L2 axis.
}
\end{figure*}

As described in the Appendix, Figure~9's black dotted curve is the current Jacobi boundary ($14\farcm24$). The light-blue annulus marks the range $8'$-$22'$ reached over the last few Gyrs. The angled blue ellipse is the \emph{projected Jacobi escape zone} of the stars most likely to leave the confines of the cluster: aligning with the local orbital tangent (not the MW center-anticenter line). Stars crossing the L1/L2 saddles are immediately sheared along the ellipse’s flanks, where the leading and lagging tails peel away.

\section{Discussion}

The SP\_Ace metallicities of our extra-tidal sample are calibrated on a solar–relative benchmark-star
scale and are derived by fitting many Fe-lines in full-spectrum synthesis,
whereas our CaT [Fe/H] values are tied to a GC-CaT calibration
that assumes a particular [$\alpha$/Fe] (and hence [Ca/Fe]) pattern and GC
metallicity scale from \citet{carretta97} \& \citet{kraft03}.  Small differences between these reference scales, together
with the sensitivity of the CaT indices to [Ca/Fe], naturally produce offsets of order 0.05-0.15~dex, such that the SP\_Ace [Fe/H] values appear
systematically more metal-rich than the CaT-based values.  Continuum-placement
uncertainties in medium-resolution spectra can introduce additional
$\sim 0.05$~dex shifts, but these are secondary compared to the underlying
scale and abundance-pattern differences. However, the CaT metallicities are calibrated to the same underlying GC scale, as discussed in \S2.4.4.

From our theoretical modeling described in \S3 and the Appendix, the P23 reported  King ``tidal'' (limiting) radius of $\simeq9\farcm8$ is fully consistent with a time-varying $r_J(t)$ that is typically larger away from pericenter, and some of our ``extra-tidal" sample overlap with J18's sample and P23's dataset (see Figure~9, near L2).

North of NGC~6569's center, and close to the dotted line between L1/L2 in Figure~9, there is an arc-like overdensity, present in both the spectroscopic targets and Gaia-PM samples. This is naturally explained as \emph{epicyclic caustics} \citep{kupper12}: stars that have recently leaked through the L1/L2 saddles and still execute near-cluster epicycles. Theory and preliminary N-body modeling predict that recent escapees cause cusp-like over-densities at $r\!\sim\!0.6$-$0.8\,r_{\rm J}$, producing a short S-shaped bridge and shell-like ridges that later shear into long collimated tails \citep[e.g.,][]{Kupper10,kupper12,FukushigeHeggie2000,Claydon2017}. Time-dependent barred/bulge tides amplify these caustics and may create a shallow envelope of potential escapees; bound (green-filled circles) and probably unbound (red-filled circles and blue-filled diamonds) stars coexist near the Roche surface (dotted curve in Figure~9).  The candidate escapees lie at $0.5$-$3\,r_{\rm J}$, exactly where such crowding is expected, consistent with systems like Pal~5 \citep{odenkirchen03,grillmair06,erkal17}. 

We plan much more extensive N-body simulations to quantify the correspondence of observed stars with suspected tidal debris in the crowded bulge. The Roche surface does not remain the same size over the entire orbit (see Appendix), and the GC is likely to leave a ``halo" of tidal debris on the edge of being bound/unbound between the tidal ``bridges" and the leading and lagging tidal tails. The \textsc{AGAMA} calculation uses the full, time-dependent, non-axisymmetric barred MW model and locates the true L1/L2 saddle points from $\nabla\Phi_{\rm eff}=0$ without constraining them to the GC-cluster line. This captures the anisotropic, phase-dependent tidal field near pericenter and apocenter, yielding a larger and more realistic ``breathing" of the Roche surface: $r_{\rm J}\!\simeq\!8\arcmin$-$10\arcmin$ at pericenter and $18\arcmin$-$22\arcmin$ at apocenter. Those values reproduce the morphology and kinematics in Figure~9: the observed neck-to-tail transition near $12\arcmin$ sits at $\sim0.8\,r_{\rm J}$ (present phase), the tail axes align with the predicted escape directions from L1/L2, and the bound/borderline/unbound classifications track the inner–plateau dispersion and the extra-tidal candidates. 

In contrast, the fast \textsc{Gala} estimate (axisymmetric $\Omega$-$\kappa$ with L1/L2 forced to lie in the direction to the MW center) damps the modulation and misorients the escape geometry, underpredicting the range seen in the data. For this reason, we adopt the \textsc{AGAMA} $r_{\rm J}(t)$ for figure overlays and membership decisions, using the \textsc{Gala} series only as a quick diagnostic. However, results from \textsc{Gala} and \textsc{AGAMA} are broadly consistent with each other once the different methodologies are taken into account.
This approach was tested with the J18 RV-selected members in the inner cluster and, at larger radii, cleanly isolates potential escapees consistent with the Figure~9 morphology. 

Over a $\geq$12 Gyr lifetime, a Kroupa/Chabrier IMF \citep{kroupa01, chabrier03, bastian10} or a more recent analysis \citep{webb24}, implies that stellar evolution alone removes $\sim40$-$50\%$ of a globular cluster’s initial mass. The remaining deficit is produced by two-body evaporation in a steady tide and by repeated bulge/bar shocks at each pericenter \citep[and references therein]{vesperini97, Kupper10}. For NGC 6569, adopting present masses $M_{\rm now}\simeq(1.7$-$2.3)\times10^5\, M_\odot$ (see Table~1), an initial-to-current ratio $k\equiv M_0/M_{\rm now}\simeq6$-$10$ implies $M_0\simeq(1$-$2.3)\times10^6,M_\odot$ and total losses of $\sim83$-$90\%$ roughly half from early stellar evolution and half from long–term tidal processes. In our bar-frame integrations, the Jacobi radius $r_J(t)$ exhibits deep minima at pericenter, coincident with minima of the critical Jacobi energy $E_{J,{\rm crit}}(t)$ (see Figure~A.1 in the Appendix), identifying epochs of strong tidal shocking, when escape through L1/L2 is maximized \citep{binney08}. 

The ASteCA analysis (\S2.2) yields a contamination index of
${\rm CI} \simeq 0.35$ in the outer regions (beyond $\sim 4\arcmin$,
near the half-light radius), indicating that our nominal ``field''
annulus remains substantially cluster-dominated in the CMD.  Although
CI does not translate directly into a ``35\% mass loss'' estimate, this
moderate value is consistent with an extended halo of marginally bound
and recently stripped stars (see Figure~9).  Together with the large, orbit-dependent Jacobi radii inferred for NGC~6569 and our chemo-dynamical evidence for ongoing stripping, the ASteCA analysis supports a scenario in which a
non-negligible fraction of the present-day stellar mass already resides
outside the classical King profile and is in transition toward the
surrounding bulge field, providing independent photometric evidence for
significant tidal dissolution.

In the Appendix, we use the extra-tidal candidates and the J18 sample to estimate the mass-loss rate, since those stars have not left the cluster boundaries (Figure~9), and 11-15 of the objects are probably unbound (Table A.3). Using the specific age/metallicity/cluster mass estimates (Table~1 and Figure~2), NGC 6569 should host $\sim 1.2-1.5\times10^2$ RGB stars, so that the J18 dataset can be considered a representative sample.
With present-day cluster mass of $M_{\rm cl}=2.3\times10^{5}\,M_\odot$, we express the inferred mass-loss rates as fractions per unit time (Appendix sections B \& C), and find: 
\noindent{
\paragraph{Central (crossing-time) mass loss estimate:}\leavevmode\\
With $\dot M_{\rm cen}\approx (1.0\text{-}1.6)\times10^{-5}\,M_\odot\,\mathrm{yr^{-1}}$,
\begin{equation}
\begin{split}
\frac{\dot M_{\rm cen}}{M_{\rm cl}}
  &= (4.35\text{-}6.96)\times10^{-11}\ \mathrm{yr^{-1}} \\
  &= 5.65^{+1.31}_{-1.30}\ \%\,\mathrm{Gyr^{-1}} .
\end{split}
\end{equation}}
\noindent{
\paragraph{Tail-based mass loss estimate:}\leavevmode\\
With $\dot M_{\rm tail}\approx (1\text{-}3)\times10^{-5}\,M_\odot\,\mathrm{yr^{-1}}$,
\begin{equation}
\begin{split}
\frac{\dot M_{\rm tail}}{M_{\rm cl}}
  &= (4.35\text{-}13.04)\times10^{-11}\ \mathrm{yr^{-1}} \\
  &= 8.70^{+4.34}_{-4.35}\ \%\,\mathrm{Gyr^{-1}} .
\end{split}
\end{equation}}

The leading sources of systematic uncertainty are: (i) the orbit sampling in the rotating–bar model (through $\Omega_\phi$ and $\Delta\phi$), (ii) the spectroscopic selection completeness $f_{\rm sel}$ used to scale number counts to mass, and (iii) the tracer–to–total mass conversion $\langle m\rangle$. These terms are propagated into the quoted ranges for $\dot M$ and $\dot M/M_{\rm cl}$, which reflect the range in $\dot M$ (from selection completeness and tracer-to-mass scaling) with $M_{\rm cl}$ held fixed. If $M_{\rm cl}$ has fractional uncertainty $\delta M/M_{\rm cl}$, then the fractional rates scale inversely: $\delta(\dot M/M)\approx(\dot M/M)\,(\delta M/M_{\rm cl})$. For reference, literature values span $M_{\rm cl}\!\simeq\!(1.7\!-\!2.3)\times10^{5}\,M_\odot$ (see Table~1 in the main text), implying a plausible fractional uncertainty of $\delta M/M_{\rm cl}\!\sim\!0.2\!-\!0.3$; adopting $M_{\rm cl}=2.3\times10^{5}\,M_\odot$ places our preferred ratios at the conservative end of this range.

We also quantified the physical bulge-field contamination with a Monte Carlo resampling test. The null hypothesis is `any apparent excess of stars beyond $\sim7'$ with NGC~6569-like kinematics and chemistry arises from random bulge draws'. This was evaluated using \bes\ bulge models convolved with our median errors in all parameters. These predict only $\langle N_{\rm field}\rangle=0.9\pm0.9$ contaminants satisfying \emph{simultaneously} the [Fe/H], $[\alpha/{\rm Fe}]$, RV, and PM cuts. Observing $N_{\rm obs}=18$ such stars occurs in $<10^{-4}$ of $10^{4}$ trials ($>3.9\sigma$). We conclude that Grade~A\,+\,B objects are \emph{overwhelmingly consistent} with tidal debris from NGC~6569 (null probability $<10^{-4}$; $\gtrsim3.9\sigma$ against the bulge–only hypothesis). Grade C \& D stars also meet the chemistry+RV+PM window and are likely genuine debris, albeit with larger individual uncertainties. 

In our observational data space, astrometry plus velocities becomes six-dimensional, \((\alpha,\;\delta, \;\varpi,\;\mu_{\alpha*},\;\mu_{\delta}, \; RV)\). Including abundances adds two more dimensions (\([\mathrm{Fe/H}], [\alpha/\mathrm{Fe}]\)), giving \(D=8\); if the parallax is negative/unreliable and omitted, we have \(D=7\). Only the A+B objects have reliable $[\alpha/\mathrm{Fe}]$, and some of that sample also lack reliable distances. All lines of evidence place the F-class stars outside the dynamical, and mostly outside the chemical, space of \emph{recent} debris: they would have had to share the RVs and chemistry of newly escaped stars to enter our final selection. The 6 Grade-F stars that have barely failed the RV- and/or chemistry selection criteria could also be included in a widened sample, but their S/N values are low. Examining the group more closely, ID nos.: 309, 336, and 345 appear too metal-rich or metal-poor, while objects 67, 195, and 253 have the correct \fedp , and are within $\pm 5$~\kms\ of our RV-range.

 We note that our red spectral range cannot measure elements beyond Fe and Ca without higher S/N (akin to the A-Grade sample). Bluer coverage is required to separate 1P/2P populations among bound and escaping stars. There is no evidence for a Fe-bimodality \citep[J18;][]{crociati23,barrera25}; \citet[J18]{barrera25} found 1P/2P stars. We might expect multiple-population studies to show kinematic differences between 1P and 2P stars \citep{carretta09,carretta10,gratton12,piotto15}, where higher-resolution spectra are also needed.

More than 80\% of our 303 AAT spectral targets lie outside NGC~6569’s RV range. Of 67 stars with $S/N>35$ or constrained $[\alpha/{\rm Fe}]$, 53 lack the RVs expected for recent escapees. Additionally, 7 stars share the cluster chemistry but have discrepant RVs; they could, in principle, be older escapees in \textsc{Gala} orbits, 6 Grade-F stars appear kinematically bound to the GC, which likely implies the GC is moving with the bulge population. Some 60 stars lie near other bulge GCs (in position and parameter space), which may provide a useful field sample for bulge chemistry/kinematics. Follow-up spectroscopy (light elements) and dynamical modeling of nearby clusters and their debris will be presented in McEwen et al.\ (2026, in preparation).

\section{Summary \& Conclusions}
\noindent
We have conducted the first wide-field, medium-resolution spectroscopic survey
of NGC~6569 that explicitly targets stars beyond the cluster’s nominal King
tidal radius \citep{valenti11}.  Combining AAOmega spectroscopy for 303 stars
with \textit{Gaia}~DR3 astrometry, BDBS photometry, and cross-matching with the
high-resolution samples of J18 and P23, we draw the following main conclusions.

We identify a population of stars at $r \simeq 7\arcmin$-$30\arcmin$ from the
center of NGC~6569 that we interpret as bona fide tidal debris: 7 Grade~A and
12 Grade~B objects with high-quality spectra and chemical-composition
membership probabilities exceeding 70\%, and 21 Grade~C and D stars with lower
S/N or only moderate probabilities.  Of this ABCD sample, only five stars have
marginally reliable \textit{Gaia}~DR3 parallaxes that might tag them as
foreground objects, yet they appear to be part of a halo of tidal debris around
NGC~6569.  The remaining 258 Grade~F stars deviate from the cluster centroid by
$>3\sigma$ in at least one of RV, PM, or chemistry and are entirely consistent
with the surrounding bulge/bar population.  

We model NGC~6569's theoretical tidal debris and potential streams with the
\textsc{gala} and \textsc{agama} codes, finding consistent orbital and Jacobi
radius predictions. Comparing the dynamically-bound cluster stars
of NGC~6569 with the surrounding field, we find that $\sim 35\%$ of the local
stellar population shares its PM.  Combining the chemo-dynamical constraints
with our dynamical analysis, we model the cluster as undergoing continuous mild
stripping at $\dot{M}/M \simeq 5.6 \pm 1.3\%~\mathrm{Gyr}^{-1}$, with an
$L_1/L_2$ boundary that ``breathes'' between $r_{\rm J} \simeq 11\arcmin$ and
$18\arcmin$ along the orbit.  The observed excess of unbound candidates in the
lagging direction points to NGC~6569 moving through a tube of its own tidal
debris; dedicated $N$-body simulations will be required to explore this
scenario more fully.

  The RV- and [Fe/H]-selected AAT stars, together with the
J18 and P23 samples, show that the outer cluster is approximately elongated
along the $L_1/L_2$ line, and that all samples curve toward the predicted
orbital path, forming a nascent S-shaped stream.  The stars graded A-D at
$r \simeq 7\arcmin$-$14\arcmin$ lie in the zone where tidal tails first emerge
in benchmark systems such as Pal~5 \citep{odenkirchen03}.  Within this region
we expect subtle signatures, including a mild axial elongation along the orbit
and a surface-density break beyond $\sim 0.6\,r_{\rm J}$; we indeed observe a
departure from circular or elliptical symmetry from $\sim 4\arcmin$ out to
$\sim 14\arcmin$ along the L1/L2 axis.  Within this annulus,
we identify a tidal bridge and the onset of the northeast lagging tail, while
the southwest leading tail is more weakly populated.  These structural
signatures are fully consistent with the ongoing mass loss inferred from our
chemo-dynamical analysis and support the picture that NGC~6569 is actively
shedding stars into the bulge field.

\paragraph{Future work:}
Deeper photometry is required along the nascent tails, as the RGB stars identified here are likely only tracers of the stellar population leaving the cluster. High-resolution spectroscopy of the Grade A \& B candidates will test whether they exhibit the Na-N-He patterns that fingerprint the second generation GC stars (first generation stars would not be chemically differentiable from the field stars). In parallel, N-body simulations tailored to NGC~6569’s present-day orbit will quantify its ongoing mass-loss rate and survival timescale in the bulge potential. Building on the successful identification of extra-tidal stars around NGC 6569, MWBest will use AAOmega @AAT to observe fields located approximately 3-4 degrees from the cluster center, along both the leading and lagging tails of NGC 6569, in order to search for extra-tidal stars at greater distances from the cluster as demonstrated in this study.
Extending this approach to the full MWBest sample will ultimately clarify the cumulative role of GC-dissolution in building the Galactic bulge.

\begin{acknowledgments}\small
We thank the anonymous referee for their insightful comments and constructive suggestions, which significantly improved the clarity and robustness of this manuscript.

\rev{Support for the pilot AAT 3.9\,m observations was provided by the Preparing for Astrophysics with LSST Program, funded by the Heising-Simons Foundation and managed by Las Cumbres Observatory. The authors gratefully acknowledge support for this project from the M.J. Murdock Charitable Trust (NS-2017321) and from NSF awards that support this collaborative research project (AST-2408324, AST-2408325, AST-2408326, AST-2408327).}

Additionally, the project used data obtained with the Dark Energy Camera (DECam), constructed by the Dark Energy Survey (DES) collaboration. Funding for DES has been provided by numerous institutions, including the U.S. Department of Energy, the National Science Foundation, and international partners. A full list of supporting agencies and collaborating institutions can be found at \url{https://www.darkenergysurvey.org/collaboration}.
Observations were conducted at Cerro Tololo Inter-American Observatory, NSF’s NOIRLab (Prop. IDs 2013A-0529, 2014A-0480; PI: M. Rich), operated by AURA under a cooperative agreement with the NSF.

This work used data from the European Space Agency (ESA) mission {\it Gaia} (\url{https://www.cosmos.esa.int/gaia}), processed by the \textit{Gaia} Data Processing and Analysis Consortium (DPAC). DPAC funding was provided by national institutions participating in the \textit{Gaia} Multilateral Agreement.

\vspace{1mm}
\facilities{AAT: 3.9m,  AAVSO, CTIO: 1.3m, 
CTIO: 4.0m, CXO.}
\rm
We are grateful to the developers of the open–source packages that
enabled this research.

\software{
    \textsc{AGAMA}\citep{agama}
    \textsc{astropy} \citep{astropy},
    \textsc{ChatGPT} \citep{openai24},
    \textsc{CLOUDY} \citep{cloudy},
    \textsc{Gala} \citep{pricewhelan17},
    \textsc{Matplotlib} \citep{hunter07},
    \textsc{NumPy} \citep{harris20},
    \textsc{Overleaf} \citep{overleaf_online_2025},
    \textsc{Scite} \citep{scite},
    \textsc{SEextractor} \citep{sextractor},
    {\tt 2dfdr} \citep[AAO; ascl:1505.015]{aao15}
}
\end{acknowledgments}




\appendix
\setcounter{table}{0}\renewcommand\thetable{A.\arabic{table}}
\setcounter{equation}{0}\renewcommand\theequation{A.\arabic{equation}}

\setcounter{table}{0}
\setcounter{equation}{0}

\renewcommand\thetable{A.\arabic{table}}
\renewcommand\theequation{A.\arabic{equation}}

\section{Modeling the Orbit, Size, Shape, and Escape from NGC 6569}
\setcounter{figure}{0}
\renewcommand{\thefigure}{\thesection.\arabic{figure}}

We modeled the \textit{Gaia} surface-density profile of NGC~6569 with an elliptical-annulus Plummer law, $\Sigma(R)=\Sigma_{0}[1+(R/a)^{2}]^{-2}+\Sigma_{\rm bg}$, adopting the outer isophotal shape ($q=b/a=0.75$, ${\rm PA}=20^{\circ}$). A weighted least-squares fit gives $a=2\farcm44^{+0.12}_{-0.15}$, $\Sigma{0}=27.6^{+3.2}_{-3.2}\ \mathrm{arcmin}^{-2}$, and $\Sigma{\rm bg}=0.138^{+0.024}_{-0.020}\ \mathrm{arcmin}^{-2}$ (68\% uncertainties from 300 Poisson bootstraps of the annular counts), with $\chi^{2}=14.9$ for 16 degrees of freedom. The model reproduces the curvature from $R\sim1\arcmin$ to $10\arcmin$ and asymptotes to the measured background by $R\sim12\arcmin\text{-}15\arcmin$, remaining consistent with the mean Jacobi scale $\bar r_{J}\simeq14\farcm2$. Compared to spherical King fits ($r_{t}\approx10\arcmin\text{-}11\arcmin$) and even shape-aware King fits ($r_{t}\approx15\arcmin$), the Plummer profile yields a lower misfit and a more faithful description of the non-truncated, prolate outer envelope.
\vspace{1cm}

\begin{deluxetable*}{lcc}
\tabletypesize{\small}
\tablecaption{Orbital periods from  \textsc{Gala}}
\tablehead{
\colhead{Quantity} & \colhead{Symbol} & \colhead{Value (Myr)}
}
\startdata
Azimuthal period (inertial frame) & $T_\phi$   & 83.19 \\
Azimuthal period (bar frame, relative) & $T_{\rm rel}$ & 53.03 \\
Radial period                         & $T_R$    & 48.60 \\
Vertical period                       & $T_Z$    & 26.60 \\
\enddata
\end{deluxetable*}

\subsection{Orbit determination with \textsc{Gala}}
We computed the present-day orbit of NGC~6569 with the \textsc{Gala} dynamics package
(\texttt{v1.10.1}: \citealt{pricewhelan17}). The cluster’s phase-space coordinates were built from
(ICRS) $(\alpha,\delta,D)=(273\fdg4121,-31\fdg8267,10.53~\mathrm{kpc})$ together with the
systemic proper motion and line-of-sight velocity adopted in this work (see Table~1).
These were transformed to a Galactocentric frame using the same solar parameters, and adopted our barred Milky Way model as implemented in \textsc{Gala} as described in \S3. We also used the same
\emph{composite potential} consisting of an axisymmetric bulge$+$disc$+$halo plus a rigidly
rotating quadrupole bar (pattern speed $\Omega_{\rm bar}$ and bar angle).
The Hamiltonian of this potential was integrated forward and backward for $1.6~\mathrm{Gyr}$ with a
fixed stepsize $\Delta t$ small compared to the vertical period ($\Delta t\simeq 0.2~\mathrm{Myr}$),
producing the time series $\{R(t),\phi(t),z(t),\mathbf{v}(t)\}$.

From the inertial frame orbit, we measured basic orbital elements:
the pericenter and apocenter radii ($R_{\rm peri},R_{\rm apo}$) from consecutive
minima and maxima of $R(t)$; the eccentricity $e=(R_{\rm apo}-R_{\rm peri})/(R_{\rm apo}+R_{\rm peri})$;
the maximum vertical excursion $|z|_{\max}$; and the specific angular momentum $L_z=R\,v_\phi$ and
energy $E=\tfrac12 v^2+\Phi_{\rm MW}$ along the track.
We defined the guiding radius $R_g$ from $L_z=R_g\,V_c(R_g)$ using the circular speed of the same
axisymmetric background. Fundamental periods were estimated as follows:
$T_R$ from the median separation of successive pericenters,
$T_Z$ from the median separation of successive $|z|$ maxima,
and the azimuthal period $T_\phi=2\pi/\langle \dot{\phi}\rangle$ from the mean angular rate.
To quantify the coupling with the rotating bar we also computed the bar-frame angle
$\phi_{\rm rot}(t)=\phi(t)-\Omega_{\rm bar} t$ and its mean drift rate
$\langle \dot{\phi}_{\rm rot}\rangle$, yielding the relative (bar-frame) period
$T_{\rm rel}=2\pi/\langle \dot{\phi}-\Omega_{\rm bar}\rangle$ shown in Figure~A.1 (left panels).

A Monte Carlo propagation obtained uncertainties on all reported orbital parameters: we drew $N\sim10^3$ realizations of the initial conditions from the
astrometric/radial-velocity covariance and repeated the integration and measurements for
each draw; quoted errors are the central $68\%$ intervals of the resulting distributions.
This procedure produces the periods and diagnostics displayed in Figure~A.1 (left panels) and supplies the
orbit used for the $r_{\rm J}(t)$ and $E_{J,\mathrm{crit}}(t)$ calculations in Figure~A.1 (right panels). Tables~A.1 \& A.2 show the numerical results, with the Jacobi radius varying over the orbit from 10\farcm95 to 18\farcm34.

\begin{figure*}
\centering
\includegraphics[trim={0 0 0 0},width=1.0\textwidth]{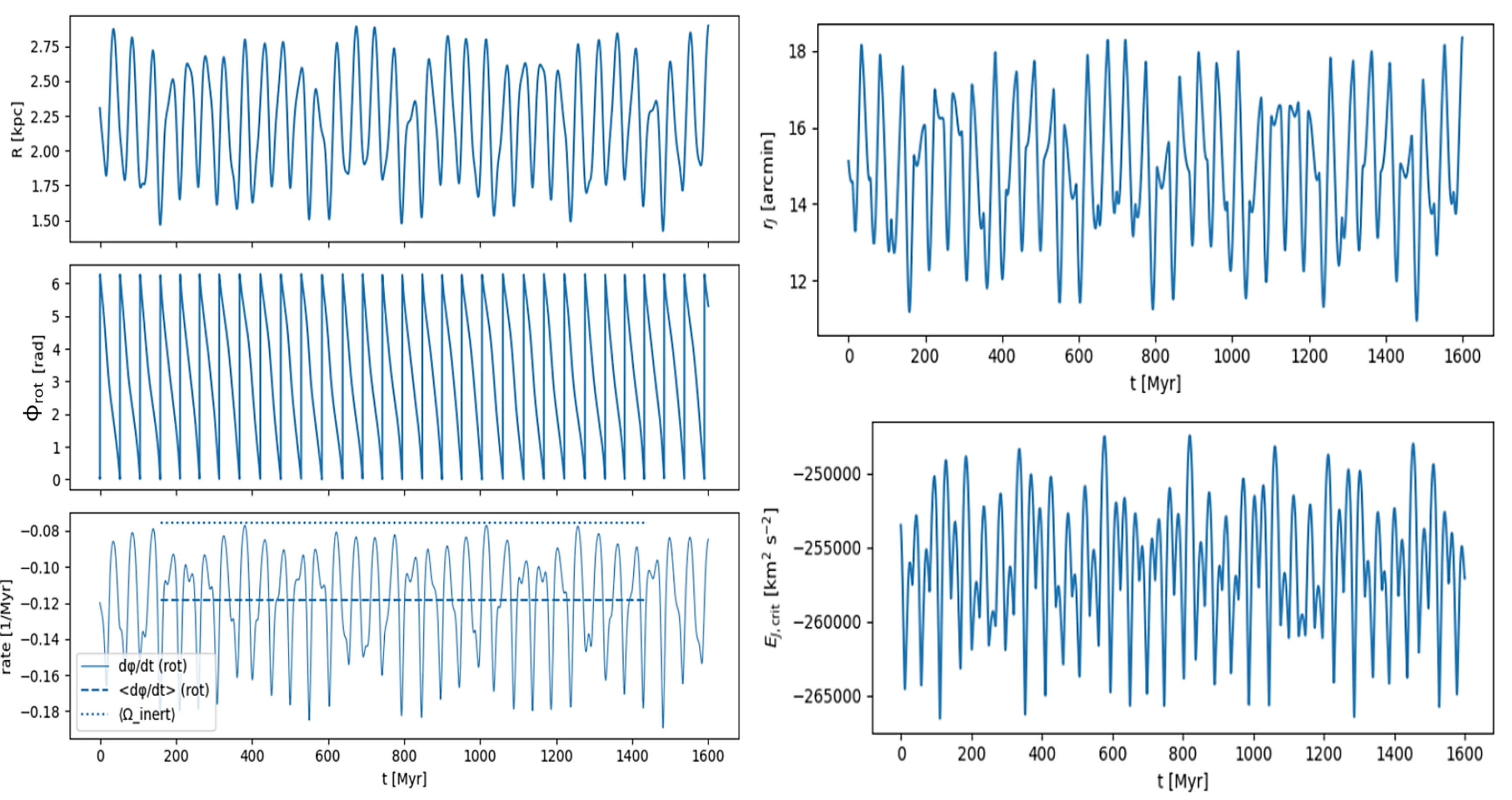}
\caption{\bf Left Column: \rm Orbital evolution of NGC~6569 in our barred–Milky Way model (integrated with \textsc{Gala}). \emph{Top}: Galactocentric cylindrical radius $R(t)$. \emph{Middle}: bar–frame azimuth $\phi_{\rm rot}(t)$. \emph{Bottom}: instantaneous bar–frame angular rate $\mathrm{d}\phi/\mathrm{d}t$ (solid), its time–average (dashed), and the inertial–frame circular frequency $\Omega$ (dotted). The listed characteristic periods ($T_\phi$ inertial, $T_{\rm rel}$ bar–frame, $T_R$, $T_Z$) are measured from the same integration. Quasi-periodic beating reflects pericenter passages and the coupling with the rotating bar. \bf Right Column: \rm Time variation of the Jacobi (Roche) radius ($r_J$) and the critical Jacobi ($E_{J,\mathrm{crit}}$) energy along the same orbit and in the same rotating–bar frame. \emph{Top}: $r_J(t)$ in arcminutes at $D=10.53$~kpc. \emph{Bottom}: $E_{J,\mathrm{crit}}(t)$, computed from the effective potential $\Phi_{\rm eff}$ evaluated at the L1/L2 saddle points. The “breathing’’ of $r_J$ (smaller near pericenter, larger near apocenter) is mirrored by the modulation of $E_{J,\mathrm{crit}}$. These curves are used to assess boundness by comparing stellar $E_J$ to $E_{J,\mathrm{crit}}(t)$.}
\end{figure*}

\begin{deluxetable*}{lcccc}
\tablecaption{Jacobi Radius Variations}
\tablehead{\\
\colhead{Quantity} & \colhead{Value} & \colhead{Units} & \colhead{Epoch} & \colhead{Notes}
}
\startdata
$r_{J,\min}$  & 0.03354 & kpc & $t=1481.200$ Myr & $10\farcm95$ \\
$r_{J,\max}$  & 0.05618 & kpc & $t=1600.000$ Myr & $18\farcm34$ \\
$\Delta t$ (between extrema) & 118.800 & Myr & \nodata & max-min separation \\
$r_{J,\max}/r_{J,\min}$ & 1.675 & \nodata & \nodata & \nodata \\
\enddata
\tablecomments{Arcminute values computed at $D=10.53$ kpc.}
\end{deluxetable*}

\subsection{\textsc{AGAMA} Modeling}
Again adopting the barred MW potential described in the main text, the present–day phase–space coordinates of NGC~6569, a distance $D=10.53~\mathrm{kpc}$, and a stellar mass $M_\star=2.3\times10^{5}\,M_\odot$, the \textsc{AGAMA} effective potential places the L1/L2 saddle points at a Jacobi (Roche) radius $r_J=$14\farcm24, i.e.\ $0.0436~\mathrm{kpc}$ ($43.6$~pc). 
This is shown below, with Eqns.[A1-3].

This exceeds the King-profile ``tidal’’ radius $r_t=6\farcm9$ \citep{valenti11} by a factor $\simeq2.06$, as expected: $r_t$ is a fit parameter of the surface-density model, whereas $r_J$ is the instantaneous limiting surface in the non-axisymmetric Galactic field, marking where leading/trailing tails launch near L1/L2. The observed neck–to–tail transition at $\sim12\arcmin$ corresponds to $\sim0.84\,r_J$ for $D=10.53$~kpc, consistent with potential–escapees dominating beyond $\sim0.5\,r_J$.

\subsection{The Roche (Jacobi) Lobe}
In the cluster’s corotating frame, the tidal (Roche) surface is the zero–velocity surface of the effective potential \citep{renaud11},
\begin{equation}
\Phi_{\rm eff}(\mathbf{x})
= \Phi_{\rm MW}(\mathbf{x}) + \Phi_{\rm cl}(\mathbf{x})
- \tfrac{1}{2}\,\Omega^2 R^2
= E_{J,{\rm crit}},
\label{eq:phi_eff}
\end{equation}
where $\Phi_{\rm MW}$ is the Milky Way potential, $\Phi_{\rm cl}$ the cluster potential, $R$ is the cylindrical Galactocentric radius, and $\Omega$ is the local circular frequency. The Lagrange points L1/L2 satisfy $\nabla\Phi_{\rm eff}=0$ and $\Phi_{\rm eff}=E_{J,{\rm crit}}$. In the axisymmetric limit,
\begin{equation}
r_{\rm J}
= \left[\frac{G\,M_{\rm cl}}{\,4\Omega^2-\kappa^2\,}\right]^{1/3},
\;
r_{\rm J}\simeq \left(\frac{G\,M_{\rm cl}}{2\,\Omega^2}\right)^{1/3}
\ \ (\kappa=\sqrt{2}\,\Omega),
\label{eq:rj_general}
\end{equation}
with $\kappa$ the epicyclic frequency \citep[e.g.,][]{binney08}. Using our adopted MW model to evaluate $(\Omega,\kappa)$ at NGC~6569 and $M_{\rm cl}=2.3\times10^{5}\,M_\odot$ yields $r_{\rm J}=0.0436~\mathrm{kpc}=43.6~\mathrm{pc}$, which subtends
\begin{equation}
\theta_{\rm J}=\frac{r_{\rm J}}{D}\times 3437.75~\mathrm{arcmin\,rad^{-1}}
\simeq 14\farcm24.
\end{equation}

\subsection{The $r_J$ Orbital Variation}
Along an eccentric inner-Galaxy orbit, $\Omega$ is largest at pericenter and smallest at apocenter, so $r_J\propto\Omega^{-2/3}$ contracts near pericenter and expands toward apocenter. For a bulge-confined path comparable to our \textsc{GALA} orbit (guiding radii $\approx$1.6-3.1 kpc), this implies a \emph{factor of} $\sim1.4$-$1.7$ modulation of $r_J$ over one revolution; pericenter values are comparable to the King $r_t$ values in Table~1.

\subsection{Boundness Probabilities and Classification}
To place the J18 RV-selected members and our extra–tidal candidates on the same footing, we evaluate boundness with a dispersion–aware escape criterion anchored to $r_J$ but \emph{not} to per–star parallaxes. For each star we form the 3D speed in the cluster rest frame from \textit{Gaia}~DR3 proper motions and line–of–sight velocities. Tangential components use
\begin{equation}
\mathbf{v}_{\rm tan}=4.74047\,\mu\,D
\quad\text{with}\quad D=10.53~{\rm kpc},
\end{equation}
after subtracting the systemic motion $(\mu_{\alpha*},\mu_\delta)=(-4.142,-7.333)\ \mathrm{mas\,yr^{-1}}$ and $v_{\rm sys}=-49.8\ \mathrm{km\,s^{-1}}$. We compare $|{\bf v}_{\rm rel}|$ to an \emph{effective} escape–speed curve from a Plummer model (scale $a=r_h/1.305$) truncated at the Jacobi surface,
\begin{equation}
v_{\rm esc,eff}(r)=\sqrt{2\,[\Phi(r_J)-\Phi(r)]},
\end{equation}
and inflate it by the cluster’s internal dispersion measured on the inner plateau, we adopt the P23 value:
\begin{equation}
\begin{split}
v_{\rm thr}(r)
  &= \sqrt{\,v_{\rm esc,eff}(r)^2+\big(n\,\sigma_{3\rm D}\big)^2\,}, \\
\sigma_{3\rm D}
  &= \sqrt{3}\,\sigma_{\rm los}, \qquad
\sigma_{\rm los}=6.5~\mathrm{km\,s^{-1}}, \qquad
n=2.5 .
\end{split}
\end{equation}

Measurement uncertainties are propagated via Monte Carlo ($N\!\sim\!10^3$ draws from the reported errors in $(\mu_{\alpha*},\mu_\delta,v_{\rm los})$ at fixed $D$), and we define the unbound probability
\begin{equation}
P=P_{\rm unbound}\equiv \Pr\!\big(|\mathbf{v}_{\rm rel}|>v_{\rm thr}(r)\big).
\end{equation}
We classify stars as \emph{bound} if $P_{\rm unbound}\le 0.2$, \emph{unbound} if $P_{\rm unbound}\ge 0.8$, and \emph{borderline} otherwise. This is our \emph{P} for each star in Table~4 of the main text.


The same procedure is applied to the AAT ABCD/RR candidates so that their $P_{\rm unbound}$ values are directly comparable to those of the J18 sample. The AGAMA Jacobi-energy calculations are used to set $r_J$ and to interpret tail kinematics, but the \emph{classification} reported here is based on the dispersion-aware $v_{\rm thr}$ criterion. 
\vspace{0.5cm}

Of the J18 sample with BDBS matches, where all stars; $D=10.53\,\mathrm{kpc}$; $v_{\rm thr}^2 = v_{\rm esc,eff}^2 + (n\,\sqrt{3}\,\sigma_{\rm los})^2$, with $\sigma_{\rm los}=6.5\,\mathrm{km\,s^{-1}}$. We find 83 of 98 stars are bound, 4 are borderline, and 11 are unbound. Cross-matching the J18 CaT giants to the P23 RV catalog yields \(83/98\) positional matches within $1\arcsec$; among these, the radial velocities agree with a median offset of $1.0\,{\rm km\,s^{-1}}$ and a median absolute difference of $1.5\,{\rm km\,s^{-1}}$. This analyis also applies to the P23 sample.

\begin{deluxetable*}{lcccc}
\tabletypesize{\footnotesize}
\tablecaption{J18 boundness breakdown (dispersion–aware method, $n=2.5$)\label{tab:j18_breakdown}}
\tablehead{
\colhead{Sample} & \colhead{Bound} & \colhead{Borderline} & \colhead{Unbound} & \colhead{Total}
}
\startdata
J18 & 83 & 4 & 11 & 98 \\
\enddata
\tablecomments{Same method and thresholds as Table~\ref{tab:abcdfrr_breakdown}.}
\end{deluxetable*}

\begin{deluxetable*}{lcccc}
\tabletypesize{\footnotesize}
\tablecaption{Boundness breakdown by grade (dispersion–aware method, $n=2.5$)\label{tab:abcdfrr_breakdown}}
\tablehead{
\colhead{Grade} & \colhead{Bound} & \colhead{Borderline} & \colhead{Unbound} & \colhead{Total}
}
\startdata
A   &  1 & 0 &   6 &   7 \\
B   &  4 & 1 &   7 &  12 \\
C   &  3 & 0 &  10 &  13 \\
D   &  3 & 0 &   5 &   8 \\
F   &  4 & 2 & 252 & 258 \\
RR  &  2 & 1 &   2 &   5 \\
\hline
TOTAL & 17 & 4 & 282 & 303 \\
\enddata
\tablecomments{Classes derived from $P_{\rm unbound}$ with thresholds: bound ($\le0.2$), unbound ($\ge0.8$), borderline (otherwise). Method uses $D=10.53$ kpc and $v_{\rm thr}^2=v_{\rm esc,eff}^2+(n\sqrt{3}\,\sigma_{\rm los})^2$ with $\sigma_{\rm los}=6.5~{\rm km\,s^{-1}}$ and $n=2.5$.}
\end{deluxetable*}

Table~4 lists all the AAT sources with $P_{\rm unbound}$ and the adopted categories. Grades ABCD candidates outside the King radius ($r_t=6\farcm9$; see Figures~7-9) appear predominantly at larger $r/r_J$ and with larger proper–motion offsets than the J18 inner–plateau members, consistent with recent or imminent escape along the leading/lagging tails.

\subsection{Comparison with Stream RV Predictions}
\label{subsec:streams}

We compared the line-of-sight velocities of the \emph{J18} \emph{unbound} subsample against two model expectations for escaping stars along the NGC~6569 stream: the ``Chen'' window ($-60<{\rm RV}<-40~{\rm km\,s^{-1}}$) and the broader ``Fardal'' window ($-65<{\rm RV}<-30~{\rm km\,s^{-1}}$, shown in the right panel of Figure~8). For the nine unbound J18 stars we find a mean and (sample) dispersion of
\begin{equation}
\langle{\rm RV}\rangle \;=\; -46.6~{\rm km\,s^{-1}}, \qquad \sigma \;=\; 10.0~{\rm km\,s^{-1}},
\end{equation}
with $5/9$ ($56\%$) falling inside the Chen interval and $9/9$ ($100\%$) inside the Fardal interval. Thus, the \emph{modal} escape velocities align more closely with the Chen prediction, while remaining fully consistent with the more permissive Fardal range.

For context, the robust (biweight) RV dispersions of the J18 subsamples are:
\begin{equation}
\begin{split}
\sigma_{\rm biwt}({\rm bound}) \simeq 4.45~{\rm km\,s^{-1}},\\
\sigma_{\rm biwt}({\rm borderline}) \simeq 4.71~{\rm km\,s^{-1}},\\
\sigma_{\rm biwt}({\rm unbound}) \simeq 9.23~{\rm km\,s^{-1}},
\end{split}
\end{equation}
i.e., the unbound group is broader by roughly a factor of two, as expected for escaping/tail stars sampling a wider range of orbital phases.

\noindent\textit{Interpretation.} The centroid and spread of the unbound velocities are centered within the Chen window and entirely contained within the Fardal bounds. This supports a picture in which recent escapers populate the L1/L2 bridges and nascent tails with RVs clustered near $-50~{\rm km\,s^{-1}}$, while a minority extend toward the edges of the broader stream velocity envelope.



\paragraph{Why does the $r_{\rm J}(t)$ modulation differ between our \textsc{Gala} and \textsc{AGAMA} runs?}
In the fast \textsc{Gala} calculation, we estimated the Jacobi (Roche) radius from the axisymmetric epicyclic
approximation,
\begin{equation}
r_{\rm J}(t)\;\approx\;\Big[\tfrac{G\,M_{\rm cl}}{\,4\Omega(t)^2-\kappa(t)^2\,}\Big]^{1/3}
\;\simeq\;\Omega(t)^{-2/3},
\label{eq:rj_axisym}
\end{equation}
and we placed the L1/L2 points on the cluster-MW Center line when evaluating $\Phi_{\rm eff}$.
With this setup (static MN$+$Hernquist$+$NFW field) the modulation amplitude is set mainly by the orbital change in
$\Omega(t)$ (or $R$): $r_{\rm J}\propto R^{2/3}$ for a nearly flat curve. For NGC~6569 the orbit used here gives
$r_{\rm J,\,min}\simeq 10\farcm95$ and $r_{\rm J,\,max}\simeq 18\farcm34$ (ratio $\simeq1.68$), fully consistent with
Eq.~(\ref{eq:rj_axisym}) and the measured $R_{\rm apo}/R_{\rm peri}$.

Our \textsc{AGAMA} calculations, instead, employ the full barred, time-dependent, non-axisymmetric Milky Way model and find the saddle points from $\nabla \Phi_{\rm eff}=0$ without constraining them to the GC line. The effective tidal tensor is therefore anisotropic, especially near pericenter, and the direction of maximum compression need not align with the L1/L2-MWC-axis. These effects, together with the bar’s pattern speed and orientation, increase the breathing of the Roche surface, yielding a larger range ($\sim8'$-$10'$ at pericenter to $\sim18'$-$22'$ at apocenter). However, the results from both treatments are consistent with each other and with the observed stellar densities.

A convenient way to capture the stronger, non-axisymmetric modulation, while retaining speed, is to use the effective tidal tensor in the rotating bar frame:
\begin{equation}
\begin{split}\\
\mathbf{T}_{\rm eff}(t)\;=\;\nabla\nabla \Phi_{\rm MW}\big(\mathbf{x}_{\rm cl},t\big)
\;-\;\Omega_{\rm bar}^2\,
\begin{pmatrix}1&0&0\\[2pt]0&1&0\\[2pt]0&0&0\end{pmatrix},
\\ \\
r_{\rm J}(t)\;\approx\;\Big[\tfrac{G\,M_{\rm cl}}{\lambda_{\max}(t)}\Big]^{1/3},
\end{split}
\label{eq:rj_tensor}
\end{equation}
where $\lambda_{\max}$ is the largest \emph{positive} eigenvalue of $\mathbf{T}_{\rm eff}$. Equation~(\ref{eq:rj_tensor}) reproduces the \textsc{AGAMA} range without root finding and naturally accounts for the bar–driven anisotropy and phase dependence. We note that the absolute scale obeys $r_{\rm J}\propto M_{\rm cl}^{1/3}$; changes in the adopted cluster mass or distance rescale $r_{\rm J}$ but do not alter the modulation ratio appreciably.

In the dynamical picture relevant for NGC 6569, the effective Jacobi radius is the direction- and phase-dependent distance to the L1/L2 saddle points in the Galactic tidal field. Using our \textsc{Gala/AGAMA} orbit in the adopted Milky Way potential, we find a mean value of $\bar r_J \simeq 14\arcmin$ ($\simeq 43.5$ pc at $D=10.53$ kpc), with a breathing range of $\sim 11\arcmin$-$18\arcmin$ over the orbit. Consistent with this, the outer isophotes are prolate (bridges between $\sim 7\arcmin$-$14\arcmin$) and the cluster appears round only inside $\sim 4\arcmin$. Because $r_J$ is neither spherical nor static, a hard truncation radius from spherical King fits (which gave $r_t \approx 10\arcmin$-$11\arcmin$ in circular annuli and $\approx 15\arcmin$ with elliptical annuli) is not a robust membership discriminator. This is precisely why calculating the escape velocity (or Jacobi energy) is preferable. Classifying stars as bound or unbound on a star-by-star basis using their 3D kinematics (proper motions plus radial velocities) at their actual $(R,\theta)$. This energy-based criterion naturally accounts for the anisotropy of the tidal boundary, its time variability along the orbit, and the observed bridge-like departures from spherical symmetry.

Compared to spherically symmetric King fits ($r_{t}\approx10\arcmin$-$11\arcmin$) and even shape-aware King fits ($r_{t}\approx15\arcmin$), the Plummer profile provides a lower misfit and a more faithful description of the non-truncated, prolate outer envelope.
The smaller $r_{\rm J}$ range from the \textsc{Gala} quicklook arises from the axisymmetric, MW Center-line approximation in Eq.~(\ref{eq:rj_axisym}); the larger \textsc{AGAMA} range reflects the full non-axisymmetric tides and the off-axis location of the Lagrange points. Both methods return broadly consistent results.
\section{Estimating the Present-Day Mass-Loss Rate of NGC\;6569}\label{sec:mdot-method}

Our estimate of the current mass-loss rate, $\dot M$, proceeds in four steps.

\paragraph{(1) Classify recent escapees.}
We tagged stars as \emph{unbound} if they lie outside the L1/L2 ``peanut'' region and are consistent with the stream kinematics: cluster-like proper motions and radial velocities following the tail trend $\mathrm{RV}(s)$, fit separately for the leading and lagging sides. Bound and borderline objects were retained for context but excluded from the mass budget.

\paragraph{(2) Convert tail arclength to time since release.}
Using the Chen+rotating-bar orbit as a time-sampler, we converted each star particle's tail arclength \emph{`s'} to a time since escape,
\begin{equation}
t_{\rm rel}(s)\;\approx\;\frac{\Delta\phi(s)}{\Omega_\phi},
\end{equation}
where $\Delta\phi(s)$ is the model azimuthal phase offset along the orbit and $\Omega_\phi$ is the azimuthal frequency at the cluster's position. This yields a time window $[t_{\min},\,t_{\max}]$ for the unbound stars in our FOV.

\paragraph{(3) Scale counts to mass (completeness-corrected).}
From the spectroscopic sample, we estimated the stellar mass in the tails within the FOV as
\begin{equation}
M_{\rm tail}\;=\;\frac{N_{\rm unbound}}{f_{\rm sel}}\;\langle m\rangle,
\end{equation}
where $f_{\rm sel}$ is the spectroscopic selection completeness (as a function of magnitude/radius) measured against the parent photometric catalog, and $\langle m\rangle$ is the mean stellar mass corresponding to the adopted isochrone and IMF over the observed magnitude range (using the same cuts applied to the data). In practice, our spectroscopic selection completeness is $f_{\rm sel}\approx$[median] with a [min-max] range across the relevant magnitude and radius, and we propagate this variation into the $\dot M$ uncertainty.

\paragraph{(4) Mass-loss rate from tails.}
The present-day mean mass-loss rate inferred from tail stars is
\begin{equation}
\dot M_{\rm tail}\;\approx\;\frac{M_{\rm tail}}{\,t_{\max}-t_{\min}\,},
\end{equation}
computed separately for the leading and lagging sides and combined as a weighted average. As a cross-check, we also estimated $M$ per epicyclic caustic and divided by the model caustic time spacing (approximately one epicyclic period), obtaining a consistent $\dot M_{\rm tail}$.

\paragraph{Assumptions and Systematics.}
This procedure assumes that (i) the \textsc{Gala}-Chen rotating-bar orbit adequately samples the recent phase-space evolution, (ii) selection completeness is well characterized, and (iii) contamination is subdominant after the RV and PM cuts. The dominant systematics are the orbit model choice (through $\Omega_\phi$ and $\Delta\phi$), the completeness correction, and the mass-to-number conversion $\langle m\rangle$.

\subsection{Present-Day Mass-Loss Rate: Central Snapshot vs.\ Tails}\label{sec:mdot-results}

\subsubsection{Central snapshot (crossing-time) estimate.}
Adopting a half-light crossing time $t_{\rm cr}\!\approx\!0.42$~Myr and a present cluster mass $M_{\rm cl}\!\approx\!2.3\times10^{5}\,M_\odot$, we use the J18 center snapshot (98 stars: 83 bound, 4 borderline, \textbf{11 unbound}).
If unbound stars remain identifiable in the center for $\sim t_{\rm cr}$, the instantaneous number-loss rate is
\begin{equation}
\dot N \;\simeq\; \frac{N_{\rm unbound,\,center}}{t_{\rm cr}}
\;\approx\; \frac{11}{0.42~{\rm Myr}}
\;\approx\; 26~{\rm stars~Myr^{-1}}.
\end{equation}
For a Kroupa-like MF (recognizing that the RGB escapees are only tracers of the mass loss), the typical escaping mass per star is $\langle m_{\rm esc}\rangle\!\sim\!0.4$-$0.6\,M_\odot$, yielding
\begin{equation}
\begin{split}
    \dot M_{\rm cen} \;\approx\; \dot N\,\langle m_{\rm esc}\rangle
\;\approx\; (10\text{-}16)\,M_\odot~{\rm Myr^{-1}} \\
=\; (1.0\text{-}1.6)\times10^{-5}\,M_\odot~{\rm yr^{-1}}.
\end{split}
\end{equation}

\subsubsection{Tail-Based Estimate.}
Using the procedure above, the tail method gives a present-day (time-averaged over the last caustic spacing) mass-loss rate of
\begin{equation}
\dot M_{\rm tail} \;\sim\; (1\text{-}3)\times10^{-5}\;M_\odot~\mathrm{yr}^{-1},
\end{equation}
where the range reflects uncertainties in completeness and in the tracer-to-total-mass conversion.

\subsubsection{Comparison and Interpretation}
The two estimates are consistent within their systematic uncertainties:
\begin{equation}
\begin{split}
\dot M_{\rm cen}\;\approx\;(1.0\text{-}1.6)\times10^{-5}\;M_\odot~\mathrm{yr}^{-1}
\\ \text{vs.}\;\;\;
\dot M_{\rm tail}\;\approx\;(1\text{-}3)\times10^{-5}\;M_\odot~\mathrm{yr}^{-1}.
\end{split}
\end{equation}

The snapshot (instantaneous, $\sim t_{\rm cr}$) and tail (residence-time, tens of Myr) estimators agree, indicating the present epoch is not atypical; small differences follow from orbital phase and from completeness and mass-to-number systematics.

\end{document}